\documentclass[twocolumn,twocolappendix]{aastex631}
\usepackage{gensymb}
\usepackage{amsmath}
\usepackage{booktabs}
\usepackage{graphicx}
\usepackage{natbib} 

\hypersetup{linkcolor=magenta,citecolor=blue,filecolor=blue,urlcolor=blue}

\graphicspath{{./}}

\begin{document}
	
\title{Distributions and Physical Properties of Molecular Clouds in the Third Galactic Quadrant: $l=[\,219\fdg75,~229\fdg75\,]$ and $b=[\,-5\fdg25,~5\fdg25\,]$}
	
\correspondingauthor{Yan Sun}
\email{yansun@pmo.ac.cn}
	
\author{Yiwei Dong}
\affiliation{Purple Mountain Observatory, Chinese Academy of Sciences, Nanjing 210023, People's Republic of China}
\affiliation{School of Astronomy and Space Science, University of Science and Technology of China, Hefei 230026, People's Republic of China}
\author[0000-0002-3904-1622]{Yan Sun}
\affiliation{Purple Mountain Observatory, Chinese Academy of Sciences, Nanjing 210023, People's Republic of China}
\author[0000-0001-5602-3306]{Ye Xu}
\affiliation{Purple Mountain Observatory, Chinese Academy of Sciences, Nanjing 210023, People's Republic of China}
\affiliation{School of Astronomy and Space Science, University of Science and Technology of China, Hefei 230026, People's Republic of China}
\author{Zehao Lin}
\affiliation{Purple Mountain Observatory, Chinese Academy of Sciences, Nanjing 210023, People's Republic of China}
\affiliation{School of Astronomy and Space Science, University of Science and Technology of China, Hefei 230026, People's Republic of China}
\author[0000-0002-7508-9615]{Shuaibo Bian}
\affiliation{Purple Mountain Observatory, Chinese Academy of Sciences, Nanjing 210023, People's Republic of China}
\affiliation{School of Astronomy and Space Science, University of Science and Technology of China, Hefei 230026, People's Republic of China}
\author{Chaojie Hao}
\affiliation{Purple Mountain Observatory, Chinese Academy of Sciences, Nanjing 210023, People's Republic of China}
\affiliation{School of Astronomy and Space Science, University of Science and Technology of China, Hefei 230026, People's Republic of China}
\author{Dejian Liu}
\affiliation{Purple Mountain Observatory, Chinese Academy of Sciences, Nanjing 210023, People's Republic of China}
\affiliation{School of Astronomy and Space Science, University of Science and Technology of China, Hefei 230026, People's Republic of China}
\author[0000-0001-7526-0120]{Yingjie Li}
\affiliation{Purple Mountain Observatory, Chinese Academy of Sciences, Nanjing 210023, People's Republic of China}
\author[0000-0001-7768-7320]{Ji Yang}
\affiliation{Purple Mountain Observatory, Chinese Academy of Sciences, Nanjing 210023, People's Republic of China}
\author[0000-0002-0197-470X]{Yang Su}
\affiliation{Purple Mountain Observatory, Chinese Academy of Sciences, Nanjing 210023, People's Republic of China}
\author[0000-0003-2418-3350]{Xin Zhou}
\affiliation{Purple Mountain Observatory, Chinese Academy of Sciences, Nanjing 210023, People's Republic of China}
\author[0000-0003-2549-7247]{Shaobo Zhang}
\affiliation{Purple Mountain Observatory, Chinese Academy of Sciences, Nanjing 210023, People's Republic of China}
\author[0000-0003-4586-7751]{Qing-Zeng Yan}
\affiliation{Purple Mountain Observatory, Chinese Academy of Sciences, Nanjing 210023, People's Republic of China}
\author[0000-0003-0849-0692]{Zhiwei Chen}
\affiliation{Purple Mountain Observatory, Chinese Academy of Sciences, Nanjing 210023, People's Republic of China}

\begin{abstract}
	We present the results of an unbiased $^{12}$CO/$^{13}$CO/C$^{18}$O ($ J=1 $--0) survey in a portion of the third Galactic quadrant (TGQ): $219\fdg75\leqslant l\leqslant229\fdg75$ and $-5\fdg25\leqslant b\leqslant5\fdg25$. The high-resolution and high-sensitivity data sets help to unravel the distributions and physical properties of the molecular clouds (MCs) in the mapped area. In the LSR velocity range from $\sim$$-1$ to $ \sim$$85~\rm km\,s^{-1} $, the molecular material successfully traces the Local, Perseus, and Outer arms. In the TGQ, the Outer arm appears to be more prominent than that in the second Galactic quadrant (SGQ), but the Perseus arm is not as conspicuous as that in the SGQ. A total of 1,502 $^{12}$CO, 570 $^{13}$CO, and 53 C$^{18}$O molecular structures are identified, spanning over $ \sim $2 and $ \sim $6 orders of magnitude in size and mass, respectively. Tight mass--radius correlations and virial parameter--mass anticorrelations are observable. Yet, it seems that no clear correlations between velocity dispersion and effective radius can be found over the full dynamic range. The vertical distribution of the MCs renders evident pictures of the Galactic warp and flare.
\end{abstract}
	
	
\section{Introduction} \label{sec:intro}
	
\begin{deluxetable*}{lcccccc}
	\tablecaption{CO Surveys of the TGQ\label{tab:intro}}
	\tablewidth{0pt}
	\tablehead{
		\colhead{Telescope} & \colhead{Transitions} & \colhead{Sky} & \colhead{Beam} &
		\colhead{Velocity}& \colhead{RMS}& \colhead{Reference}\\
		\colhead{ } & \colhead{($J=1$--0)} & \colhead{Coverage } & \colhead{Size} &
		\colhead{Resolution}& \colhead{Noise}& \colhead{ }\\
		\colhead{} & \colhead{ } & \colhead{ } & \colhead{ } &
		\colhead{($\rm km\,s^{-1}$)}& \colhead{($ T\rm_{MB} $)(K)}& \colhead{ }	
		}
	\startdata
		{CfA 1.2 m}&   $^{12}$CO  &All {Galactic plane} &	$ 8\farcm8$ &	1.3&0.1--0.4&	(1)\\
		{}&     &$194\degree\leqslant l\leqslant270\degree $,&	&	&	& (2)\\
		&&$-4\degree\leqslant b\leqslant1\degree $&&&&\\
		{NANTEN 4 m}& 	$^{12}$CO&	$220\degree\leqslant l\leqslant60\degree $,&	$ 2\farcm6 $ &0.65	&	0.4&(3) \\
		&&$|b| \leqslant 10\degree  $&&&&\\
		{Nagoya 4 m}& 	$^{13}$CO&	$208\degree\leqslant l\leqslant230\degree $,&$ 2\farcm7 $  &	0.1&	0.5&	(4)\\
		&&$-20\degree\leqslant b\leqslant10\degree $&&&&\\
		{{Mopra} 22 m}&$^{12}$CO/$^{13}$CO/&	$ l\sim224\fdg4 $,&$\sim$$ 38\arcsec$ &	$\sim$0.1&0.3--0.9&	(5)\\
		&C$^{18}$O/C$^{17}$O&$ b\sim-0\fdg6$&&&&\\
		{Nobeyama 45 m}& $^{12}$CO/$^{13}$CO/&$ 	10\degree\leqslant l \leqslant50\degree $; $ 	198\degree\leqslant l \leqslant236\degree $, &$\sim$$ 14\arcsec$ &	$\sim$1.3&	0.6/0.3\tablenotemark{\footnotesize a}&	(6)\\
		&C$^{18}$O&$|b| \leqslant 1\degree $&&&&\\
		{ARO 12 m}& $^{12}$CO/$^{13}$CO&$ 220\degree< l<240\degree $, & $\sim$$55\arcsec$&	      0.26/0.65\tablenotemark{\footnotesize b}	&0.3--1.3&(7)\\
		&&$-2\fdg5< b<0\degree $&&&&\\
		PMO 13.7-m& $^{12}$CO/$^{13}$CO/&$10\degree\lesssim l\lesssim230\degree$,  &	$\sim$$ 50\arcsec$ &$\sim$0.16&0.5/0.3\tablenotemark{\footnotesize c}&(8)\\
		{}&C$^{18}$O &$|b| \leqslant5\fdg25$& &	&	&
	\enddata
	\tablerefs{(1) \citet{Dame_1987, Dame_2001}; (2)	\citet{May_1993,May_1997}; (3) \citet{Mizuno_2004}; (4) \citet{Kim_2004}; (5) \citet{Olmi_2016}; (6) \citet{Umemoto_2017}; (7) \citet{Benedettini_2020,Benedettini_2021}; and (8) \citet{Su_2019}.}
	\tablenotetext{\footnotesize a}{0.6 K for $^{12}$CO, and 0.3 K for $^{13}$CO and C$^{18}$O.}
	\tablenotetext{\footnotesize b}{0.26 and 0.65 $\rm km\,s^{-1}$ for two backend filter banks, respectively.
 	}
	\tablenotetext{\footnotesize c}{The present work. 0.5 K for $^{12}$CO, and 0.3 K for $^{13}$CO and C$^{18}$O.}
\end{deluxetable*}
	
CO has been the most {commonly used} tracer of the molecular gas since its first detection \citep{Wilson_1970}, and numerous CO line surveys have been conducted over the decades \citep[and references therein]{Heyer_2015}. Unlike the first quadrant of the Galaxy, the third Galactic quadrant (TGQ) has neither been extensively observed nor widely discussed. {Table~\ref{tab:intro} summarizes the CO surveys of the TGQ. Using the CfA 1.2 m survey data, \citet{Rice_2016} and \citet{Miville_2017} produced their individual all-Galaxy molecular cloud (MC) catalogs by applying different cloud detection algorithms. Combining the $^{12}$CO data of \citet{Mizuno_2004} with {\it Herschel} data, \citet{Elia_2013} studied the star formation activity in the TGQ. More recently, \citet{Benedettini_2020,Benedettini_2021} have studied the large-scale molecular material distributions and properties in this region by using the Forgotten Quadrant Survey's $^{12}$CO and $^{13}$CO (1--0) line data, which were conducted by the Arizona Radio Observatory (ARO) $12\rm~m$ antenna. Although our knowledge of the molecular interstellar medium has improved, new data, with the combination of multitracers, high-sensitivity, high-resolution, and wide Galactic coverage, can offer more accurate and comprehensive insights into the TGQ.}
	
{Such data are currently being provided by the ongoing Milky Way Imaging Scroll Painting (MWISP) project, an unbiased large-scale $^{12}$CO, $^{13}$CO, and C$^{18}$O ($ J=1 $--0) survey conducted by the Purple Mountain Observatory (PMO) 13.7 m telescope~\citep{Su_2019}. To date, the Galactic range of $l=[\,219\fdg75,~ 229\fdg75\,]$ and $b=[\,-5\fdg25,~5\fdg25\,]$ (hereafter, the G220 region) has been fully mapped by the MWISP project. {This study aims to investigate the essential information of the spatial distribution and physical properties of MCs in the G220 region.} The entire G220 region covers a total of 105 deg$^2$, which contains several well-separated spiral arm features (the Local, Perseus, and Outer arms) along the line of sight.} {The high-quality MWISP data can provide a more comprehensive cloud sample than previous studies, especially for the Outer arm. Therefore, we can expect to reveal new features of the molecular gas distribution within this region.}

	
In this paper, we present the results of the MWISP $^{12}$CO/$^{13}$CO/C$^{18}$O ($ J=1 $--0) survey toward the G220 region ($J=1$--0 may be omitted hereafter, for simplicity). In Section \ref{sec:data}, we summarize the observations, data reduction, and cloud identification. In Section \ref{subsec:large_scale}, we present the {large-scale structures traced by $^{12}$CO, $^{13}$CO, and C$^{18}$O data}. Estimates of and statistics for the physical properties are exhibited from Sections \ref{subsec:dist} to \ref{subsec:relations}. Section \ref{subsec:Example} describes five {particular} MCs. Spatial distributions of the MCs are shown in Section \ref{sec:gas_distribution}. A summary can be found in Section \ref{sec:sum}.
	
\section{Data}\label{sec:data}
\subsection{Observations and Data Reduction}\label{subsec:Obversion}
\begin{figure*}[ht!]
	\centering
	\includegraphics[width=0.9\textwidth]{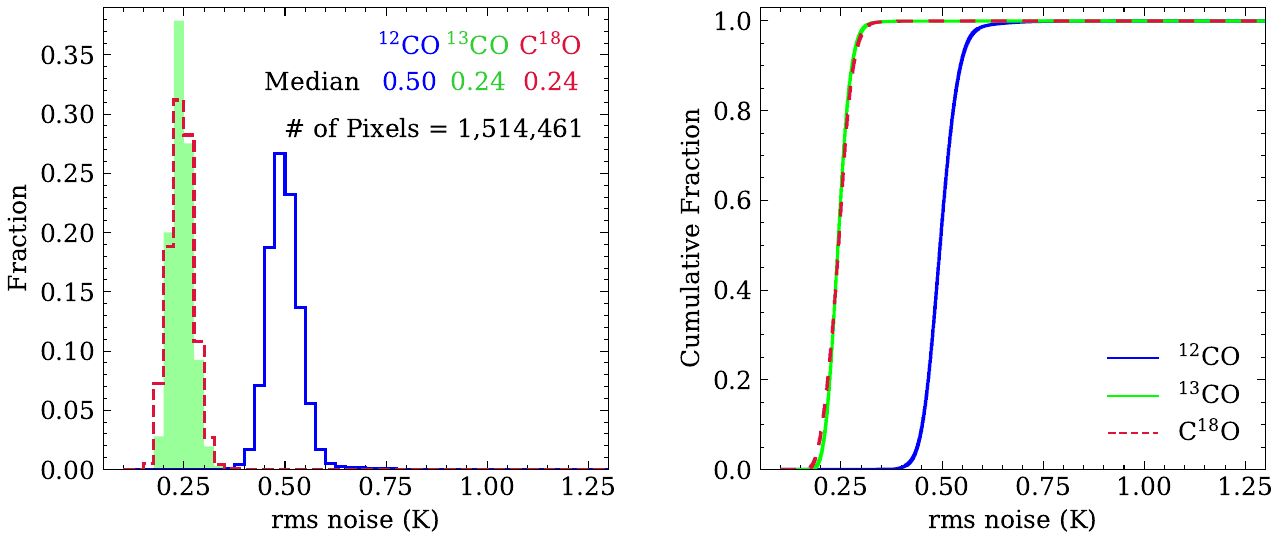}
	\caption{{Left}: histograms of the rms noise levels ($\sigma$) for $^{12}$CO (blue), $^{13}$CO (green), and C$^{18}$O (red). The median values of $\sigma$ and the total number of pixels are marked in the top right corner. {Right}: cumulative distributions of $\sigma$. \label{fig:rms}}
\end{figure*}
	
The observations were conducted from 2016 December to 2021 April by using the PMO $13.7~\rm m$ millimeter-wavelength telescope in Delingha, China. The telescope, with a half-power beam width of $\sim$$ 50\arcsec$ at $115\rm~GHz$, operates with a $ 3 \times 3 $ multibeam sideband separation receiver named the Superconducting Spectroscopic Array Receiver \citep{Yang_2008, Shan_2012}. The backend of the receiver system comprises 18 fast Fourier transform spectrometers, each configured with $1\rm~GHz$ bandwidth and 16,384 spectral channels, which yield a velocity coverage of $\sim$2,600$\rm~km\,s^{-1}$, and yield velocity resolutions of $0.159\rm~km\,s^{-1}$ at $115\rm~GHz$ and $0.166\rm~km\,s^{-1}$ at $110\rm~GHz$. The $^{12}$CO line is observed at the upper sideband, while the $^{13}$CO and C$^{18}$O lines are observed at the lower sideband simultaneously.
	
The G220 region is divided into 420 small tiles for the on-the-fly (OTF) mapping, each with a size of $30\arcmin\times30\arcmin$. The CLASS software from the GILDAS package\footnote{\url{https://www.iram.fr/IRAMFR/GILDAS}} \citep{GILDAS} is used for data reduction. The OTF data are resampled into $30\arcsec\times30\arcsec$ pixels and converted to three-dimensional (3D) FITS cubes after a first-order baseline is applied to the spectra. The velocity channels are not smoothed. The data in the FITS cubes are on the main-beam brightness temperature ($T_{\rm MB}$) scale, which is converted from the antenna temperature ($ T^\ast_{\rm A} $) with the {typical} beam efficiencies of $ \sim $0.51 for $^{13}$CO and C$^{18}$O and $ \sim $0.46 for $^{12}$CO (see the status report of the telescope)\footnote{\url{http://www.radioast.nsdc.cn/mwisp.php}}. Adding the 420 tiles together, we create the final large mosaic, which contains 1,514,461 spectra for each CO isotopolog. Please also refer to \citet{Su_2019} for more details about the observations and data reduction. Figure \ref{fig:rms} shows the normalized distributions of rms noise levels for $^{12}$CO (blue), $^{13}$CO (green), and C$^{18}$O (red). The median rms noise values are $ \sim $$0.50\rm~K$ for $^{12}$CO at a channel width of $ \sim $$0.16\rm~km\,s^{-1}$, and $ \sim $$0.24\rm~K$ for $^{13}$CO and C$^{18}$O at a channel width of $ \sim $$0.17\rm~km\,s^{-1}$.
	
\subsection{Cloud Identification and Bad Channel Cleaning}\label{subsec:identify}
An MC is popularly defined as a set of contiguous voxels in the position-position-velocity (PPV) space, with the brightness temperatures above a threshold \citep{Heyer_2015}. Based on this definition, we employ an approach developed by \citet{Yan_2020} to identify the molecular structures from the $^{12}$CO, $^{13}$CO, and C$^{18}$O data. {Note that the identified $^{12}$CO structures are often referred to as $^{12}$CO clouds or MCs.} The technique is based on the density-based spatial clustering of applications with noise (DBSCAN) algorithm\footnote{\url{https://scikit-learn.org/stable/modules/generated/sklearn.cluster.DBSCAN.html}} \citep{Ester_1996}, which contains three free parameters: $\epsilon$ and MinPts define the connection property of the voxels in the PPV space, while $ T_{\rm cutoff}$ defines the boundary isosurface of the structures \citep{Yan_2020}. A judicious option is $\epsilon=1$, $\rm MinPts=4$, and $ T_{\rm cutoff}=2\sigma $, ensuring most of the significant emission being identified without omitting the faint but real signals. \citet{Yan_2020} also recommend the usage of post-selection criteria to exclude the noise {and bad channel} contamination as much as possible. A sample will be rejected from the raw catalog if it does not meet the following criteria: (1) number of voxels $\geqslant16$; (2) peak main-beam brightness temperature $\geqslant 5\sigma$; (3) projection area contains more than one compact $2\times2$ {pixels ($ 1\arcmin\times1\arcmin $ area)}; and (4) the number of velocity channels $ \geqslant3 $. 
	
\begin{figure*}[ht!]
	\centering
	\includegraphics[width=0.9\textwidth]{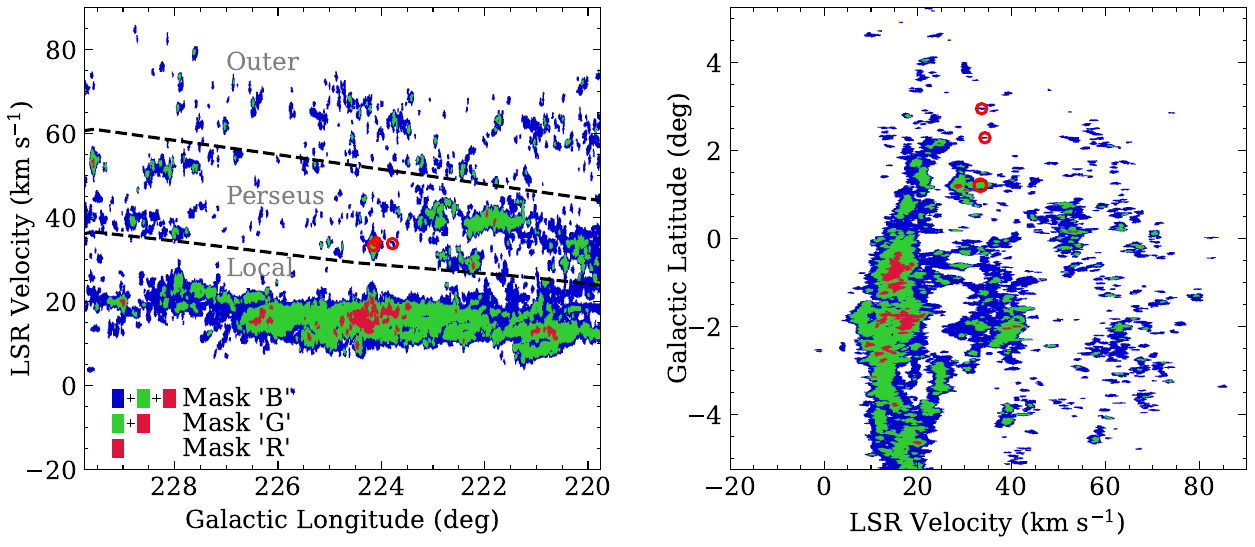}
	\caption{Longitude--velocity ({left}) and latitude--velocity ({right}) maps of the ``cleaned'' CO emission, integrated over a Galactic latitude range from $-5\fdg25$ to $5\fdg25$ and a Galactic longitude range from $219\fdg75$ to $229\fdg75$, respectively. The maps {only contain} emission from the samples in our catalog. Mask ``B,'' ``G,'' and ``R'' indicate the regions where $^{12}$CO, $^{13}$CO, and C$^{18}$O emission exists, respectively. The three contaminated MCs are marked as the red circles. The dashed lines sketch out the {boundaries} of each velocity layer. \label{fig:pvmap}}
\end{figure*}

{We then apply the DBSCAN signal identification results to mask the raw data cubes. However, we see that the bad channels centered at $\sim$$ 34~\rm km\,s^{-1} $ with a typical width of four to five channels (see Figure \ref{fig:raw_pvmap} in Appendix \ref{sec:raw_data}) cannot be completely excluded by the post-selection criteria. Therefore, we visually inspect all identified structures based on their spatial and velocity features in the PPV space. Fortunately, we find that the bad channels contaminate only three real MCs (MWISP G224.153$+$1.211, MWISP G223.784$+$2.950, and MWISP G224.097$+$2.290, marked as the red circles in Figures \ref{fig:pvmap}, \ref{fig:maskm0}, and \ref{fig:raw_pvmap}), while the majority of the bad channels simply produce spurious samples that contain only bad channels without the presence of real signal and show no signs of hierarchical structures. To get rid of the effects of these bad channels, we further exclude the spurious samples to make the final catalog (see Table \ref{tab:catalog} in Appendix \ref{sec:catalog}), and ignore the three contaminated $^{12}$CO clouds as well as their associated $^{13}$CO structure (MWISP G224.166$+$1.218) when estimating physical properties.}

\begin{deluxetable}{lcrccr}
	\tabcolsep=2.5pt
	\tablecaption{Number of Samples \label{tab:sampleNum}}
	\tablehead{
		\colhead{Layer} & \colhead{Emission} & \colhead{Full} & \colhead{Flawed} &
		\colhead{Sliced} & \colhead{Counted} \\
		\colhead{Name} & \colhead{Line} & \colhead{Sample} & \colhead{Sample} &
		\colhead{Sample} &
		\colhead{Sample}}
	\decimalcolnumbers
	\startdata
		{ Local Arm }&   $^{12}$CO   &  830$ \quad$&0&6&824$ \quad$\\
{		}& $^{13}$CO   &  419$ \quad$&0&1&418$ \quad$\\
{		}& C$^{18}$O   &  48$ \quad$&0&0&48$ \quad$\\
\hline
{ Perseus Arm }& $^{12}$CO   &  404$ \quad$&3&6&395$ \quad$\\
{          }&   $^{13}$CO   &  112$ \quad$&1&0&111$ \quad$\\
{          }&   C$^{18}$O   &  5$ \quad$&0&0&5$ \quad$\\
\hline
{ Outer Arm }&  $^{12}$CO   &  268$ \quad$&0&4&264$ \quad$\\
{          }&   $^{13}$CO  & 39$ \quad$&0&1&38$ \quad$
	\enddata
	\tablecomments{``Full Sample'' in Column 3 includes all confident samples in our catalog. Column 4, ``Flawed Sample'', gives the number of samples contaminated by the bad channels (see Section \ref{subsec:identify}). Note that one of the contaminated $^{12}$CO clouds contains a $^{13}$CO structure that is also treated as a contaminated sample. Column 5, ``Sliced Sample'', lists the number of samples that touch the mapped borders. These structures are truncated by the edges of the G220 region, leaving less than half of the complete molecular structures. By subtracting Columns 4 and 5 from Column 3, we get Column 6 (``Counted Sample''), the number of samples that are used for the analysis of physical properties (see Sections \ref{subsubsec:clouds_para}, \ref{subsubsec:ratios} and \ref{subsec:relations}).}
\end{deluxetable}
	
\section{Results}\label{sec:results}
\subsection{Large-scale Structures Traced by $^{12}{\rm CO}$, $^{13}{\rm CO}$, and ${\rm C}^{18}{\rm O}$}\label{subsec:large_scale}
The number of samples is listed in Table \ref{tab:sampleNum}. In total, we detected 1,502 $^{12}$CO, 570 $^{13}$CO, and 53 C$^{18}$O molecular structures. One can see that the number of the $^{12}$CO structures of the Local arm is more than twice that of the Perseus arm and more than three times that of the Outer arm. Figure \ref{fig:pvmap} displays the ``cleaned'' $l$--$v$ and $b$--$v$ diagrams, i.e., {only containing emission from the confident molecular structures identified by this study. Following} \citet{Sun_2020}, we define the mask ``B,'' ``G,'' and ``R'' regions as the regions where $^{12}$CO, $^{13}$CO, and C$^{18}$O emission exists, respectively. Note that the mask ``R'' region contains $^{12}$CO and $^{13}$CO emission, and the mask ``G'' region contains $^{12}$CO emission as well. According to this definition, the red mask ``R'' region in Figure \ref{fig:pvmap} is included in the {green} mask ``G'' region, and the latter is further included in the {blue} mask ``B'' region.
	
{We find that the molecular gas emission is located in the LSR velocity ($v\rm_{\scriptscriptstyle LSR}$) range from} $\sim$$-1$ to $ \sim$$85~\rm km\,s^{-1} $ {in Figure \ref{fig:pvmap}.} The dashed lines are defined as the boundaries that have the {same velocity separation} to the adjacent arm centers {\citep[refer to Figure 3 of][]{Reid_2019}}. {It is intriguing that the new high-quality MWISP data successfully trace not only the two nearby spiral arms, but also the distant Outer arm. However, from the $b$--$v$ map, it seems that our current Galactic latitude coverage is still very limited in revealing a complete view of the Local arm.} {Also note that we} do not subdivide the interarm regions {since their contribution to} the total emission is quite small. 
	
\begin{figure*}[ht!]
	\plotone{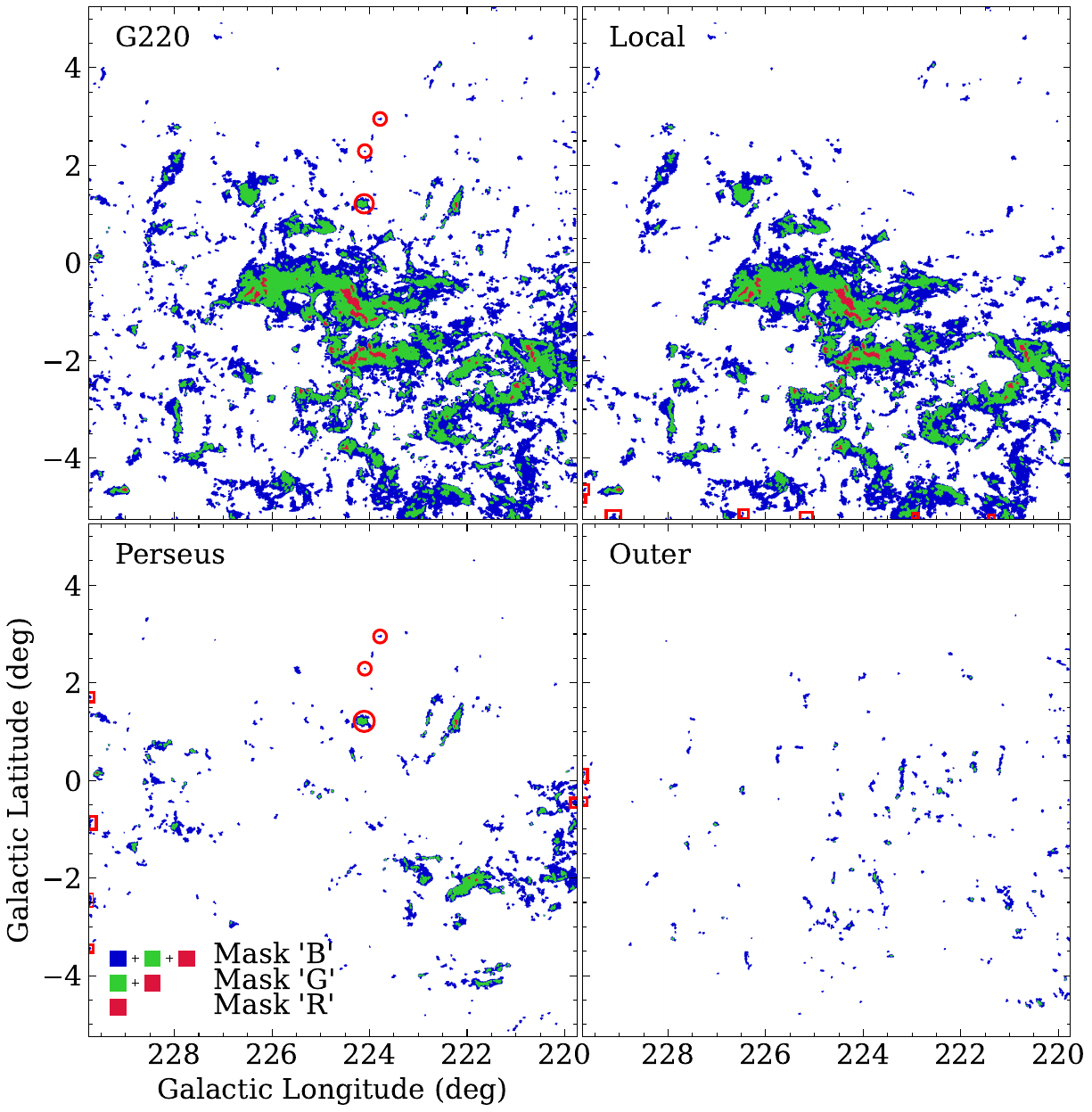}
	\caption{Mask regions of the entire G220 region ({top-left}) and the Local ({top-right}), Perseus ({bottom-left}), and Outer ({bottom-right}) arms. The definitions of the mask ``B,'' ``G,'' and ``R'' regions are the same as in Figure \ref{fig:pvmap}. {The red circles and rectangles mark the structures that are contaminated by bad channels and touch the borders of the G220 region, respectively. These structures will be ignored in the statistical analysis.} \label{fig:maskm0}}
\end{figure*}

The {mask maps of integrated intensity of the whole} G220 region {and the three arm features therein} are shown in Figure \ref{fig:maskm0}, where the mask ``B,'' ``G,'' and ``R'' regions are defined the same as in Figure \ref{fig:pvmap}. {We see that both $^{12}$CO and $^{13}$CO emission are detected in all velocity layers, while C$^{18}$O emission is only present in the two more nearby spiral arms at our current sensitivities. The decrease in flux completeness with the increasing distance may largely account for the nondetection of C$^{18}$O emission in the Outer arm.} 
	
\begin{deluxetable}{lcrD}
	\tablecaption{Layers and Mask Regions\label{tab:pixels}}
	\tablewidth{0pt}
	\tablehead{
		\colhead{Layer} & \colhead{Mask} & \colhead{Pixel} & \multicolumn{2}{c}{Pixel} \\
		\colhead{Name} & \colhead{Region} & \colhead{Number} & \multicolumn{2}{c}{Percentage}}
	\decimals
	\startdata
		{ Local Arm }&   B   &  244,729  & 16.2$\%$ \\
		{}& G   &  83,004  &5.5$\%$	 \\
		{}& R   &  4,176  & 0.3$\%$	 \\
		\hline
		{ Perseus Arm }&   B   &  41,999  &  2.8$\%$ \\
		{          }&   G   &  8,712 & 0.6$\%$	 \\
		{ }&R&123&0.008\% \\
		\hline
		{ Outer Arm }&   B   &  13,066 	 &  0.9$\%$ \\
		{          }&   G   & 1,071	&0.07$\%$
	\enddata
	\tablecomments{Pixel percentage {is defined as} the ratio of the pixel number in each mask region to the total number of pixels in the G220 region, i.e., 1,514,461.}
\end{deluxetable}
	
{The pixel numbers and pixel percentages (the ratios of pixel numbers to the total pixel number in the entire map, i.e., 1,514,461) of the mask regions} are listed in Table \ref{tab:pixels}. Obviously, {only a small fraction ($\lesssim$16.2\%) of the mapped pixels show CO emission, and the pixel percentages decrease rapidly from the nearest arm to the farthest arm. However, CO emission seems to occupy most of the mapped area in the inner Galaxy, e.g., in the Galactic longitude of $ l=[25\fdg8$, $49\fdg7 $] \citep[see Figure 6 of][]{Su_2019}. In the second Galactic quadrant (SGQ), \citet{Du_2017} and \citet{Sun_2020} also reported pixel percentages of $\lesssim$26.2\% and $\lesssim$35.7\% in the G140 region ($139\fdg75\leqslant l\leqslant149\fdg25$) and the G130 region ($129\fdg75\leqslant l\leqslant140\fdg25$), respectively. Since all of these studies used the MWISP survey data with uniform quality and employed very similar statistical methods, the differences strongly imply that, on average, the molecular gas becomes sparser when moving from the inner to the outer Galaxy.} In addition, {as expected,} the number of pixels in the mask ``B'' region is the largest, while that in the mask ``R'' region is the smallest in each spiral arm layer.
	
\begin{figure}[ht!]
	\centering
	\includegraphics[scale=0.67]{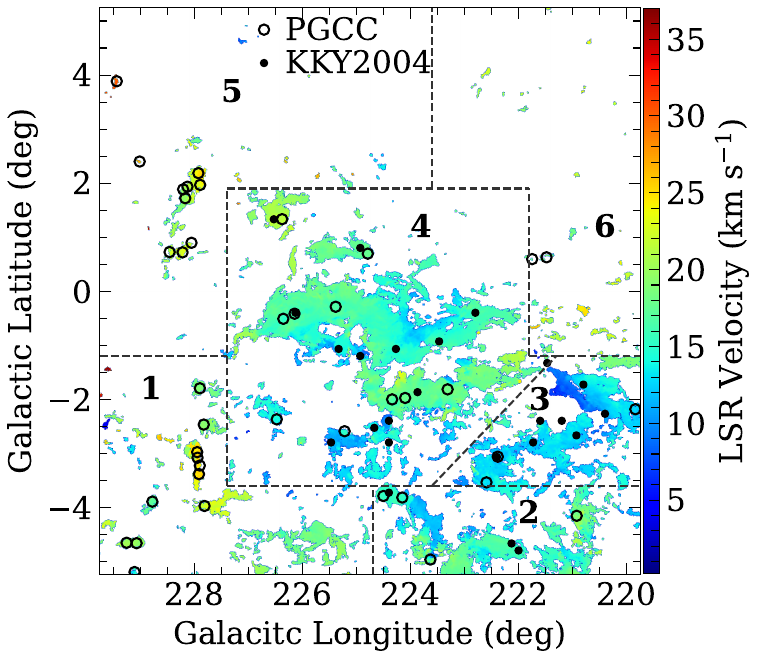}
	\caption{The velocity distribution of $^{12}$CO in the Local arm. The dashed lines indicate the boundaries of the six regions where different distances were adopted. The cold clumps (empty circles) of the \citet{PLANCK_2016} and the $^{13}$CO clouds (filled circles) of \citet{Kim_2004} are used as references for determining the distances in the six regions.\label{fig:Locm1}}
\end{figure}
	
\subsection{Distances to MCs}\label{subsec:dist}
	
The distances to the MCs in the outer Galactic plane are widely determined by assuming a rotation curve, the so-called kinematic method. However, this method is limited by large uncertainties for nearby clouds. Therefore, for the MCs in the Local arm, we adopt the distances from the literature that are mainly determined by the dust extinction and photometric methods. These methods generally provide more reliable distance measurements than the kinematic method. The \citet{PLANCK_2016} estimated the distances of the cold clumps (the empty circles in Figure \ref{fig:Locm1}) using the extinction method. \citet{Kim_2004} zonally determined the distances of the $^{13}$CO clouds (the filled circles in Figure \ref{fig:Locm1}) in the Canis Major region. Accordingly, the Local arm can be divided into six regions by the dashed lines in Figure \ref{fig:Locm1}. {For simplicity, we assign a constant distance for the MCs within each region. The adopted distance and the number of MCs in each region} are given in Table \ref{tab:Locdist}. Region 1 is tentatively further subdivided into two cases: a distance of $\rm 0.2~kpc$ is adopted for the relatively nearby clouds with low $v\rm_{\scriptscriptstyle LSR}$ \citep[the uncertainty-weighted average distance of the cold clumps in this region;][]{PLANCK_2016}, and a distance of $\rm 2~kpc$ is adopted for the relatively distant clouds with high $v\rm_{\scriptscriptstyle LSR}$ \citep{Montillaud_2015}. Also using the uncertainty-weighted average method, the distances to the MCs in regions 2, 5, and 6 are determined as 0.6, 0.3, and $\rm 0.7~kpc$, respectively. Referring to \citet{Kim_2004}, we assign 1 and $\rm 1.15~kpc$ \citep[the photometric distance to the CMa OB1 stellar association;][]{Claria_1974} to the MCs in regions 3 and 4, respectively.
	
Considering that very few references are available, we derive the kinematic distances for the MCs in the Perseus and Outer arms. The universal rotation curve of \citet[two-parameter version]{Persic_1996} and parameter values from \citet[model fit A5]{Reid_2019} are adopted here. In the G220 region, MWISP G229.573$+$0.146 is the only MC with high-accuracy distance measurements \citep[a parallax distance of $4.59^{+0.27}_{-0.24}\rm~kpc $ is measured by][and a kinematic distance of $ \sim$$ 4.9\rm~kpc $ is obtained and adopted by us]{Choi_2014}. However, the arm assignment for MWISP G229.573$+$0.146 remains controversial. In this study, we still follow the results of \citet{Reid_2019} to assign this source to the Perseus arm. And it is worth noting that, {more recently, \citet{Xu_2023}} consider{ed} this source to belong to the Outer arm.
	
\begin{deluxetable}{lrrrrrrr}
	\tablecaption{Regions and Distances in the Local Arm\label{tab:Locdist}}
	\tablewidth{0pt}
	\tablehead{
		\colhead{Region} & \multicolumn2c{1} & \colhead{2} & \colhead{3} &
		\colhead{4}& \colhead{5}& \colhead{6}
		}
	\startdata
		{Reference}& (1) & $ \!\! $(2) &(1)&	(3)&	(3)&	(1)&	(1)\\
		{Dist. (kpc)\tablenotemark{\footnotesize a}}& 0.2&	$ \!\! $2&	0.6&	1 &	1.15&	0.3&	0.7\\
		{$ \# $ of Clouds\tablenotemark{\footnotesize b}}& 118&	$ \!\! $19&	136&	92 &	335&	60&	70
	\enddata
	\tablerefs{(1) \citet{PLANCK_2016}; (2)	\citet{Montillaud_2015}; and (3) \citet{Kim_2004}.}
	\tablenotetext{\footnotesize a}{Heliocentric distance to each region.}
	\tablenotetext{\footnotesize b}{Number of $^{12}$CO clouds in each region.}
\end{deluxetable}
	
\begin{deluxetable*}{lcrrrrrrrcrrrr}
	\tablecaption{Statistics of Excitation Temperatures, H$_2$ Column Densities, Masses, and Areas\label{tab:Tex_N_H2}}
	\tablewidth{0pt}
	\tablehead{
		\colhead{Layer} & \colhead{Mask} &\multicolumn{5}{c}{$ T_{\rm ex}~\left(\rm K\right)$}& \multicolumn{5}{c}{$\log\!\left[N_{\rm H_2}~\left(\rm{cm}^{-2}\right)\right]$}& \colhead{Mass}&\colhead{Area}\\
		\cmidrule(l{4pt}r{4pt}){3-7}
		\cmidrule(l{4pt}r{4pt}){8-12}
		\colhead{Name}&\colhead{Region}&\colhead{Min.} & \colhead{Max.} &
		\colhead{Med.}& \colhead{Avg.}& \colhead{S.D.}&\colhead{Min.} & \colhead{Max.} &
		\colhead{Med.}& \colhead{Avg.}& \colhead{S.D.}&\colhead{($M_\odot $)}&\colhead{(pc$^2$)}
		}
	\decimalcolnumbers
	\startdata
		{Local Arm}&  B$ \ $&	3.7 &	36.5 &	5.9 &	6.9&	3.03$ \ $&	19.4 &	22.3 &	20.8 &	20.7&	0.59	&$ \ $	$ 1.5\times10^5 $ &5,242\\
		{}&G$ \ $&	3.9 &	36.5 	&9.1 &	9.9	&3.47$ \ $&19.4 &	22.7 &	20.8&	20.8&	0.62	&$ \ $	$ 7.1\times10^4 $&2,020\\
		{}& R$ \ $&	6.1 &	32.9 &	12.8 &	14.9&	5.31$ \ $&	20.5 &	22.7 &	21.4 &	21.4&	0.45	&$ \ $	$ 1.0\times10^4 $&111\\
		\hline
		{Perseus Arm}&  B$ \ $&	3.6 &	31.8 &	5.2 &	6.0&	2.45$ \ $&19.4 &	22.2 &	20.5 &	20.5&	0.55	&$ \ $	$ 1.7\times10^5 $&11,245\\
		{}&G$ \ $&	3.9 &	31.8 &	8.5 &	9.5	&3.46$ \ $&	19.5 &	22.5 &	20.6 &	20.7&	0.62	&$ \ $	$ 6.5\times10^4 $&2,247\\
		{}& R$ \ $&	8.6 &	28.9 &	11.0 &	14.6	&5.48$ \ $&20.6 &	22.1 &	21.3 &	21.3&	0.37	&$ \ $	$ 2.0\times10^3 $&30\\
		\hline
		{Outer Arm}&  B$ \ $&	3.7 &	10.9 &	4.9 &	5.1	&0.94$ \ $&	19.4&	21.6 &	20.3 &	20.3&	0.45	&$ \ $	$ 8.3\times10^4 $&11,911\\
		{}& G$ \ $&	4.1 &	10.7 &	6.6 &	6.7&	1.07$ \ $&19.6 &	21.6 &	20.4 &	20.4&	0.43	&$ \ $	$7.6\times10^3$ &948\\
		\enddata
	\tablecomments{``Min.'', ``Max.'', ``Med.'', ``Avg.'', and ``S.D.'' are the abbreviations for minimum, maximum, median, average, and standard deviation.}
\end{deluxetable*}
	
\subsection{Statistics of Physical Properties}\label{subsec:Statistics}
The derivation of the physical properties using the $^{12}$CO, $^{13}$CO, and C$^{18}$O lines is described in detail in Appendix \ref{sec:derivation}. We remind the reader that the calculation of $ T_{\rm ex} $ is based on the assumption that the beam-filling factor is equal to unity. However, the actual values are often less than unity for the majority of MCs. Using the simulation results of \citet[Equation 9 therein]{Yan_2021_filling_factor}, the median filling factors of the $^{12}$CO structures in the Local, Perseus, and Outer arms are estimated to be about 0.66, 0.65, and 0.64, respectively. Hence, the excitation temperatures derived in this study represent the lower limits. 
	
We should also note that $N{\rm_{H_2}}$ traced by {the $^{12}$CO alone} is obtained using the X-factor method, while those traced by $^{13}$CO and C$^{18}$O are derived using the LTE method. We know that $X_{\rm CO}$ may vary in changing environments, and early works suggest that $X_{\rm CO}$ increases from the Galactic center to the outer disk, due to the decreasing metallicity \citep{Brand_1995,Sodroski_1995,Strong_2004}. 
{Applying the method of \citet{Barnes_2015, Barnes_2018} to the $^{12}$CO clouds harboring $^{13}$CO structures, we estimate the mean values of $ X\rm_{CO} $ to be about 1.4, 1.6, and 2.2 $\times 10^{20}~\rm cm^{-2}\left(K~km/s\right)^{-1}$ for the Local, Perseus, and Outer arms, respectively. 
These values do not significantly deviate from $X_{\rm CO}=2.0 \times 10^{20}~\rm cm^{-2}\left(K~km/s\right)^{-1}$ \citep{Bolatto_2013}, so the commonly used value is assumed throughout this study.}
	
\subsubsection{Image-based Statistics$\rm\,: $ $ T_{\rm ex} $, $N{\rm_{H_2}}$}\label{subsubsec:Global}
\begin{figure*}[ht!]
	\includegraphics[width=1\textwidth]{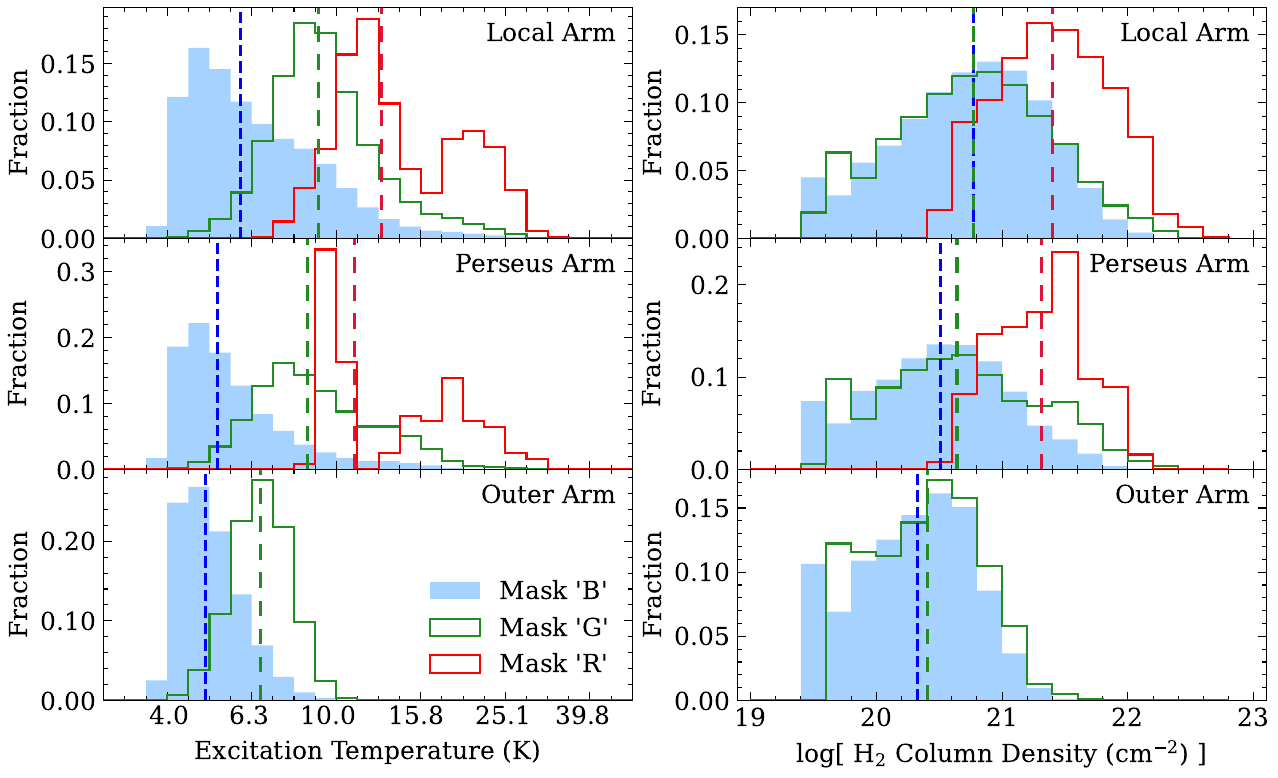}
	\caption{Histograms of the excitation temperatures ({left}) and H$_2$ column densities ({right}). Top to bottom: histograms of the Local, Perseus, and the Outer arms. The mask ``B,'' ``G,'' and ``R'' regions are colored in blue, green, and red, respectively. {The vertical dashed lines in the corresponding colors indicate the median values.} \label{fig:Tex_N_H2}}
\end{figure*}

The {pixel-by-pixel} statistics of the excitation temperatures ($ T_{\rm ex} $) and H$_2$ column densities ($N{\rm_{H_2}}$) are listed in Table \ref{tab:Tex_N_H2} and plotted as histograms in Figure \ref{fig:Tex_N_H2}. {We find that} the excitation temperatures and H$_2$ column densities in the mask ``B,'' ``G,'' and ``R'' regions increase {progressively}. {In addition, the two nearby arms show very similar $ T_{\rm ex} $ and $N{\rm_{H_2}}$ values, while they show higher $ T_{\rm ex} $ and $N{\rm_{H_2}}$ values than the Outer arm. {The different levels of beam dilution effects may partially contribute to this trend.} When comparing to other Galactic environments, we find that the values for both properties in the G220 region are lower than those in the G130 region \citep{Sun_2020}, suggesting the relatively low-excitation and low-density environment in the G220 region.}
	
	
Furthermore, we notice the bimodal distributions of the excitation temperatures in the mask ``R'' regions in both the Local arm and the Perseus arm, and the lower peaks appear to dominate the distributions. For the Local arm, the two components are regarded to be contributed by two types of MCs. The ``cold'' component with a peak at $ T_{\rm ex}\sim 12\rm~K$ is contributed by quiescent clouds or clouds in the early stages of star formation, while the ``warm'' component with a peak at $T_{\rm ex}\sim 22\rm~K$ is related to MCs containing active star-forming activity~\citep{Lin_2021}. For the Perseus arm, the two components may also be due to two different types of MCs.
	
\subsubsection{Sample-Based Statistics}\label{subsubsec:clouds_para}
\begin{figure*}[ht!]
	\centering
	\includegraphics[width=1\textwidth]{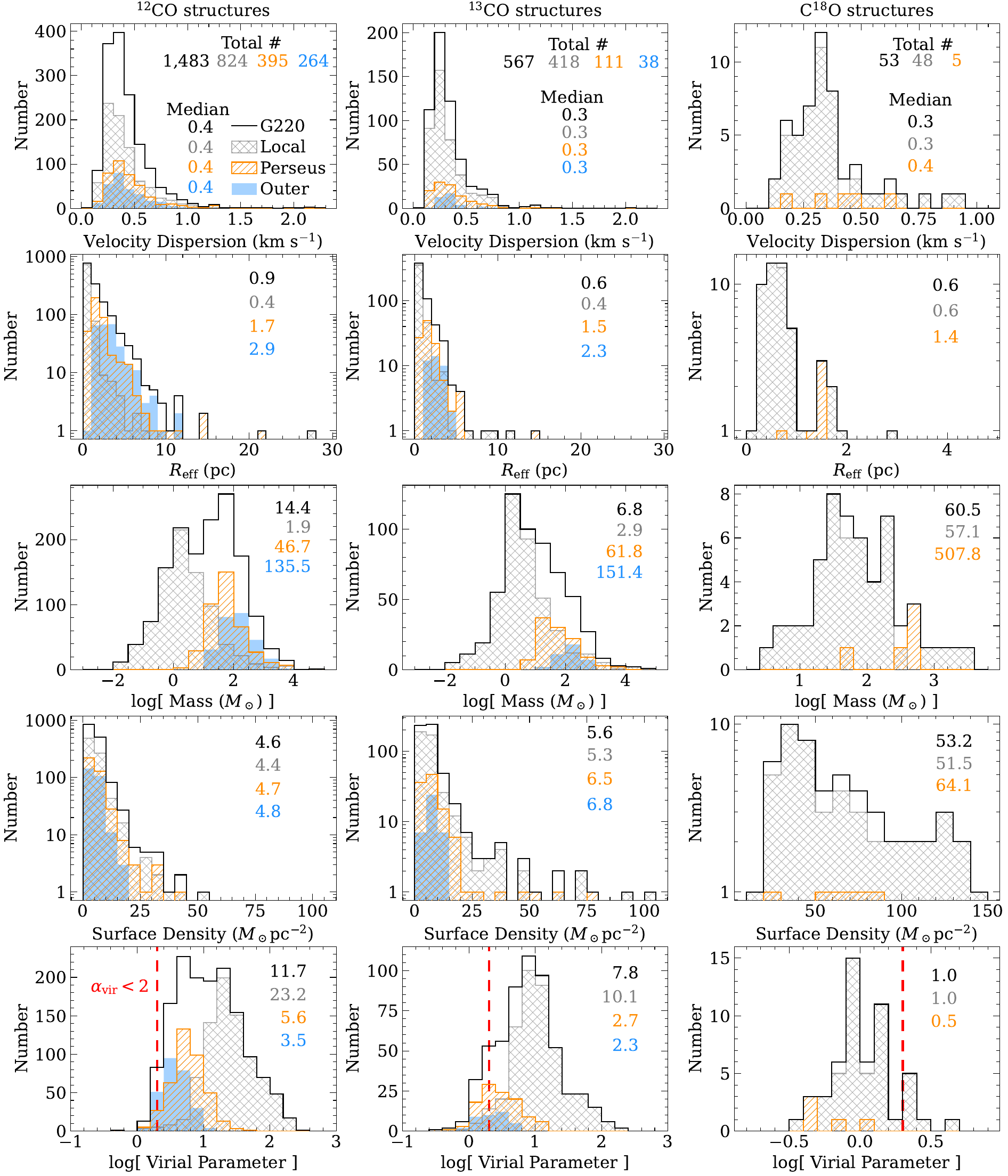}
	\caption{Histograms of the derived physical properties of {the $^{12}$CO ({left}), $^{13}$CO ({center}) and C$^{18}$O ({right}) structures. Distributions for the structures in the whole G220 region and the Local, Perseus, and Outer arms are shown in black, gray, orange, and blue histograms, respectively.} {Top to bottom}: histograms of the velocity dispersion, effective radius, mass, mass surface density, and virial parameter. The median values are labeled in each panel. The total numbers of molecular structures included in the statistics are marked in the top panels. The vertical dashed lines in the bottom panels denote the loci of $\alpha \rm_{vir}=2$. \label{fig:para}}
\end{figure*}

{The physical properties of the $^{12}$CO, $^{13}$CO, and C$^{18}$O structures within the three spiral arms are presented and compared in this section, including the velocity dispersion ($\sigma_v$), effective radius ($R\rm_{eff}$), mass ($M$), mass surface density ($\Sigma$), and virial parameter ($\alpha \rm_{vir}$). {Please refer to Appendix \ref{sec:derivation} for their definitions}. Figure \ref{fig:para} exhibits the histograms of their distributions. The numbers of structures are labeled in the top panels. The median values for each property are also labeled in each panel. Note that the scale range of the x-axis is fixed for each property of the $^{12}$CO and $^{13}$CO structures due to their similar dynamic ranges, while it is wider than that for the C$^{18}$O structures.} Generally speaking, the outskirts of {an individual MC} are traced by $^{12}$CO, {the moderately dense structures are traced by $^{13}$CO, and the most dense structures are traced by C$^{18}$O. At our current sensitivities, the vast majority of the $^{12}$CO clouds do not harbor $^{13}$CO and C$^{18}$O structures. And most of the $^{13}$CO structures do not harbor C$^{18}$O structures either (see Section \ref{subsubsec:ratios} for details). Therefore, the physical properties derived by the three isotopologs may have different distributions.} {Besides, plots of the mean and median values of the physical properties at different Galactocentric radii are displayed in Figure \ref{fig:vary} in Appendix \ref{sec:vary}.}

The velocity dispersions {in Figure \ref{fig:para}} show narrow distributions that span over $\sim$0.1--2.2$~\rm km\,s^{-1}$, $\sim$0.1--2.0$~\rm km\,s^{-1}$, and $\sim$0.1--0.9$~\rm km\,s^{-1}$ for the $^{12}$CO, $^{13}$CO, and C$^{18}$O structures, respectively. 
{For} each individual CO isotopolog, the median values of $\sigma_v$ appear to be the same across the Galactic disk ({around} $ \sim $0.4, 0.3, and 0.3$~\rm km\,s^{-1}$ for the $^{12}$CO, $^{13}$CO, and C$^{18}$O structures, respectively). {For the $^{12}$CO structures,} the median value of $\sigma_v$ {is} much lower than those shown by the data from the CfA survey (e.g., {$\sim$ 2.8 $\rm km\,s^{-1}$}; \citealt{Miville_2017}) and from the MWISP survey with different decomposition algorithms (e.g., {$\sim$ 1.1 $\rm km\,s^{-1}$}; \citealt{Ma_2021}). However, {it is} similar to those shown by \citet{Sun_2021} from their results for the MWISP data with the same DBSCAN cloud decomposition method.
Note that the velocity resolution of the data, the cloud decomposition algorithm, and the local environment of the interstellar medium may all affect the statistics of $\sigma_v$.

\begin{figure*}[ht!]
	\centering
	\includegraphics[width=0.9\textwidth]{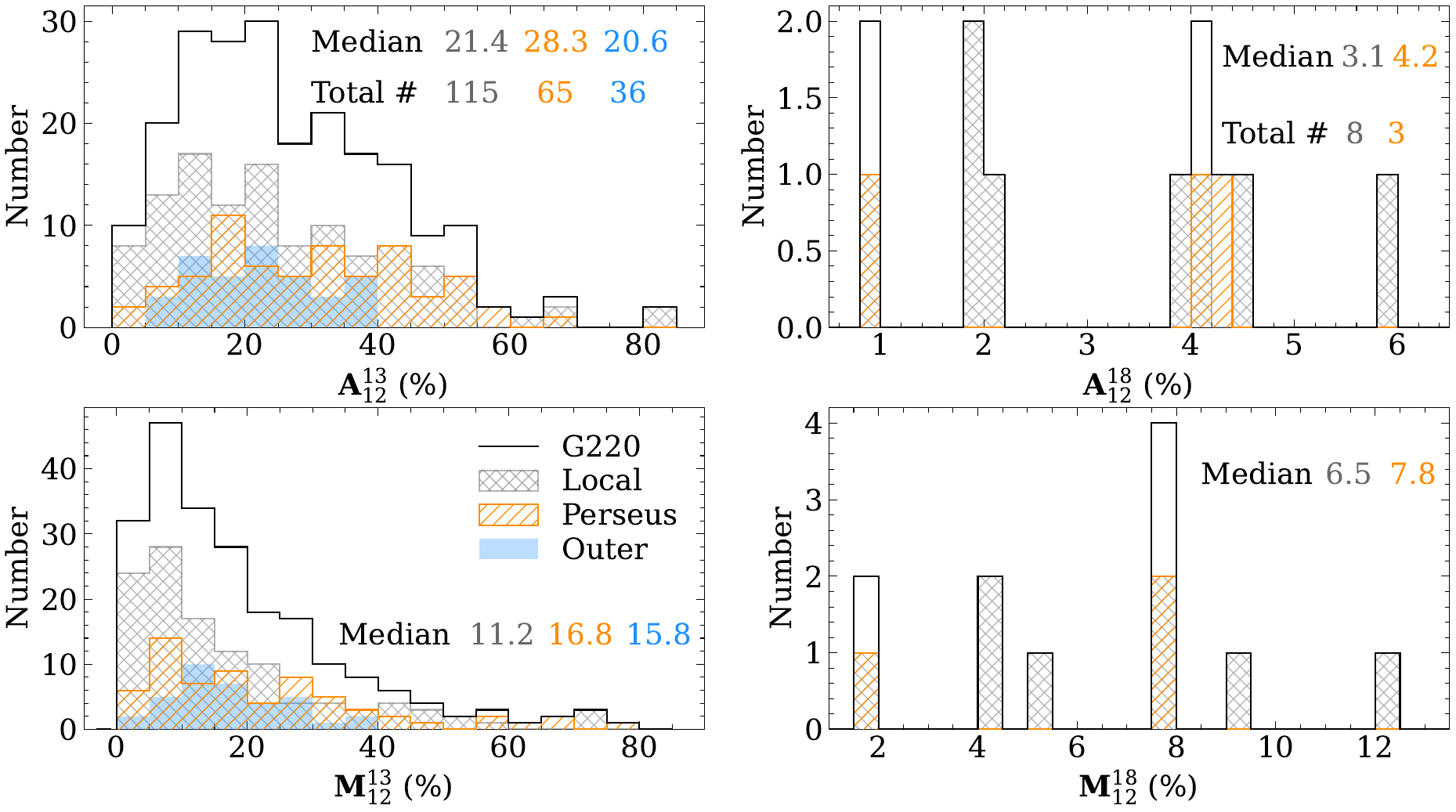}
	\caption{Histograms of $\mathbf{A}^{13}_{12}$ ({top-left}), $\mathbf{A}^{18}_{12}$ ({top-right}), $\mathbf{M}^{13}_{12}$ ({bottom-left}), and $\mathbf{M}^{18}_{12}$ ({bottom-right}) of MCs in the whole G220 region (black) and the Local (gray), Perseus (orange), and Outer (blue) arms. Please refer to Section \ref{subsubsec:ratios} for the definitions of the area and mass ratios. {The total numbers of structures included in the statistics and the median values for the ratios are marked.}\label{fig:ratio}}
\end{figure*}

{The cloud sizes and masses of our sample span about 2 and 6 orders of magnitude, respectively.} 
As expected, the dynamic ranges of $R\rm_{eff}$ and $M$\textbf{, visible in Figure \ref{fig:para},} show the highest, medium, and lowest values for the $^{12}$CO, $^{13}$CO, and C$^{18}$O structures, respectively.
{Due to the improved sensitivity and resolution of the MWISP data, we detect a large number of small and/or faint structures with subparsec radii and subsolar masses that were largely} missed or identified as parts of the large MCs in previous studies. {The relatively small and/or faint $^{12}$CO and $^{13}$CO structures typically do not contain C$^{18}$O structures, possibly due to the C$^{18}$O structures suffering more severe beam dilution effects.} 
When compared to the {median} values that are based on the CfA data within the G220 region, such as \citet[{$\sim $17 pc and $ \sim $$5\times 10^3~M_\odot $}]{Miville_2017} and \citet[{$\sim $17 pc and $ \sim $$8\times 10^3~M_\odot $}]{Rice_2016}, our $^{12}$CO samples show the much smaller median values for both $R\rm_{eff}$ and $M$. However, our $^{12}$CO samples again show similar median $R\rm_{eff}$ and $M$ values to those of \citet[see Tables 6 and 7 therein]{Sun_2021}.

{In the entire G220 region, the total masses of the $^{12}$CO, $^{13}$CO, and C$^{18}$O structures are $ 4.0\times10^5 $, $ 1.4\times10^5 $, and $ 1.2\times10^4~M_\odot $, respectively~(listed in Column 13 of Table \ref{tab:Tex_N_H2}).} The Local, Perseus, and Outer arms account for $ \sim$$38\% $, $ \sim$$41\% $, and $ \sim$$21\% $ of the total $^{12}$CO-traced mass in the G220 region, respectively. It seems that the two nearby spiral arms contribute a similar portion of the total molecular gas mass and are more prominent than the Outer arm. Nevertheless, the Perseus arm is much more prominent than the other arms in the G130 region~\citep{Sun_2020}, with the Local, Perseus, and Outer arms accounting for $\sim$20\%, $\sim$70\%, and $\sim$10\% of the total $^{12}$CO-traced mass, respectively. Besides, both our $^{12}$CO- and $^{13}$CO-traced masses in the Outer arm are two times larger than those of the G130 region. These {trends} follow the scenario traced by the young stellar population \citep{Carraro_2005, Moitinho_2006, Vaquez_2008} and \ion{H}{1} gas \citep{Koo_2017}---the Perseus arm {becomes less prominent}, while the Outer arm seems to grow, not lessen, from the SGQ to the TGQ.

{The mass surface densities of the structures traced by different CO isotopologs vary greatly---by $\sim$2 orders of magnitude. As expected, the $^{12}$CO structures tracing the relatively diffuse portion of the MCs have the smallest median values of $\Sigma$, while the C$^{18}$O structures tracing the relatively dense portion of the MCs have the largest median values of $\Sigma$.
{For} each individual CO isotopolog, the median values of $\Sigma$ {are almost constant across different Galactocentric radii (around $\sim$4.6, 5.6, and 53.2$~M_\odot\,\rm pc ^{-2}$ for $^{12}$CO, $^{13}$CO, and C$^{18}$O structures, respectively)}.} The nearly constant median value of $\Sigma\approx 5~M_\odot\,\rm pc ^{-2}$ for $^{12}$CO clouds at $R_{\rm GC}\gtrsim10\rm~kpc$ is also reported in previous studies~\citep[e.g.,][]{Miville_2017, Sun_2021}.
	
\begin{figure*}[ht!]
	\centering
	\gridline{\fig{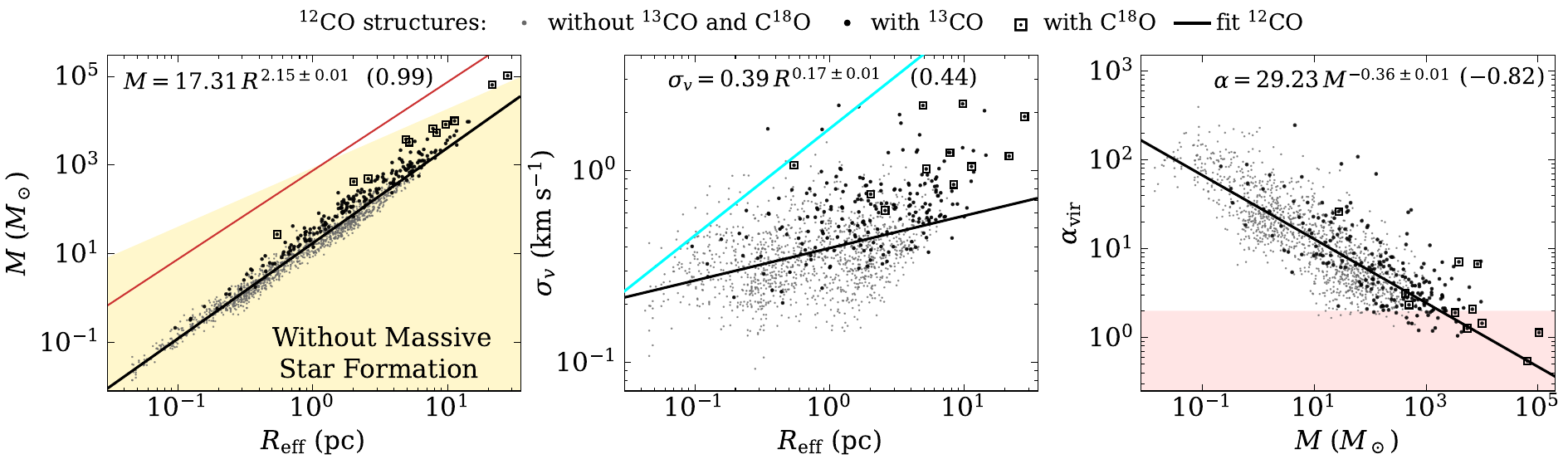}{1\textwidth}{}}
	\vspace{-0.8cm}
	\gridline{\fig{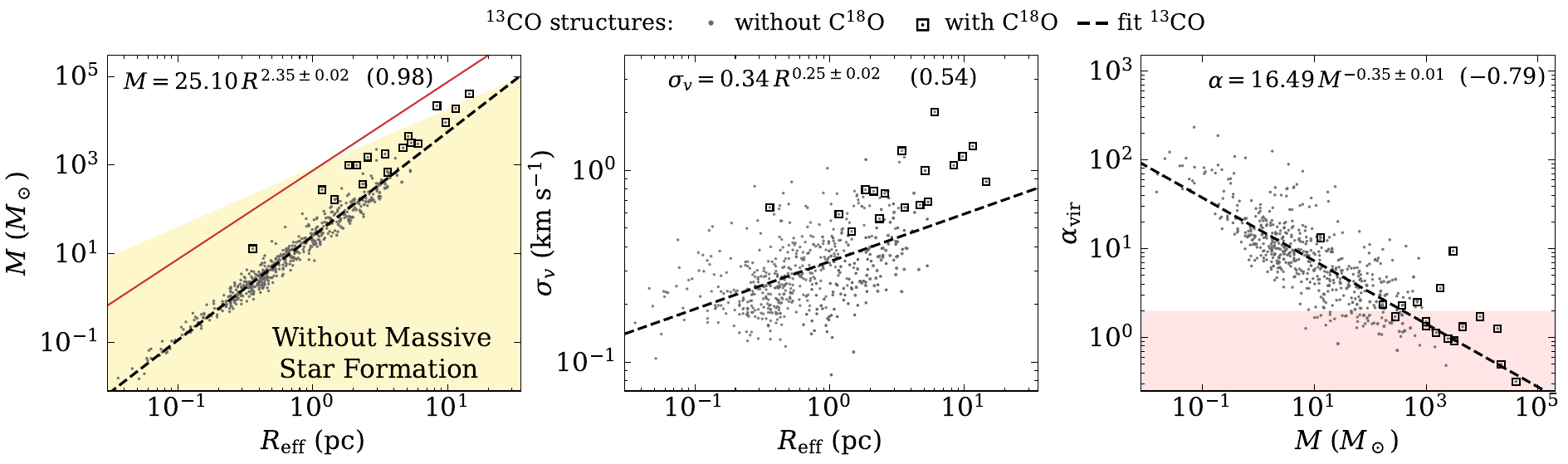}{1\textwidth}{}}
	\vspace{-0.8cm}
	\gridline{\fig{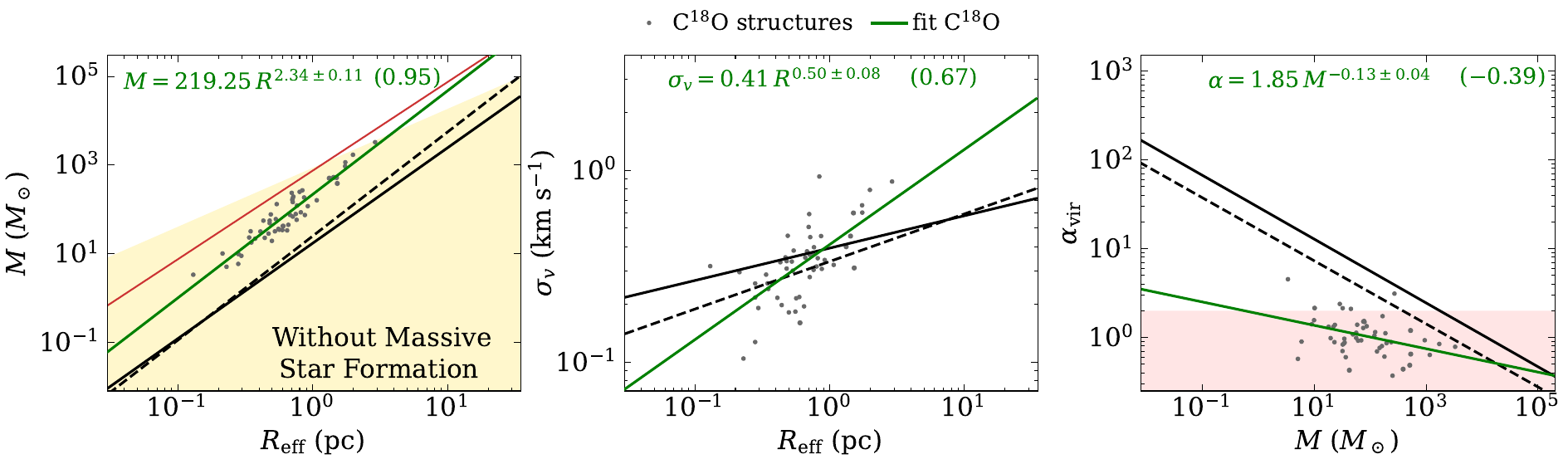}{1\textwidth}{}}
	\vspace{-0.8cm}
	\caption{Relations between physical properties of the molecular structures traced by different CO isotopologs. {{Top row}}: most $^{12}$CO structures are plotted as small dots; those containing $^{13}$CO structures are plotted as bigger dots in darker colors; and those further containing C$^{18}$O structures are {enclosed in black} squares. {{Middle row}}: similar to the top panels, but for $^{13}$CO structures. {{Bottom row}}: for C$^{18}$O structures. {{Left column}}: mass vs. effective radius diagram. The yellow shadings, $ M(r) \leqslant 870\, M_\odot\left(R/\rm pc\right)^{1.33} $, highlight the regions where the clouds fail to form high-mass stars \citep{Kauffmann_2010a,Kauffmann_2010b, Kauffmann_2013,Kauffmann_2010c}. The solid red lines are the empirical lower limit of the surface density required for massive star formation \citep[$ \Sigma=0.05~\rm g\,cm^{-2} $,][]{Urquhart_20131, Urquhart_20132}. {{Middle column}}: velocity dispersion vs. effective radius plot. The first Larson relation revealed by the data from \citet{Solomon_1987} is also drawn with the cyan line for comparison. {{Right column}}: virial parameter as a function of mass. The {pink} shadings indicate the regions of $ \alpha \rm_{vir} <2$. The black solid, black dashed, and green solid lines delineate the log--log linear least-squares fitting relations between the physical properties of the $^{12}$CO, $^{13}$CO, and C$^{18}$O structures, respectively. The fitting results are also annotated in each graph, with the Pearson correlation coefficients marked in parentheses. \label{fig:relation}}
\end{figure*}
	
The virial parameter describes the balance between the kinetic and gravitational energies of the MCs. The vertical dashed lines in the last row of Figure \ref{fig:para} demarcate the gravitationally bound structures ($\alpha \rm_{vir}<2$) from the unbound structures ($\alpha \rm_{vir}\geqslant2$; \citealt{Kauffmann_2013}). We find that only a very small fraction of $^{12}$CO structures appear to be bound, with the percentages of $\sim$$0.5 \%$, $\sim$$4.1 \%$, and $\sim$$8.0\%$ for the Local, Perseus, and Outer arms, respectively. And there are $\sim$$3.1 \%$, $\sim$$32.4 \%$, and $\sim$$42.1\%$ gravitationally bound $^{13}$CO structures in the Local, Perseus, and Outer arms, respectively. Unlike the $^{12}$CO and $^{13}$CO structures, the C$^{18}$O structures are more likely to be bound, i.e., $ \sim$$86.8 \%$ of the C$^{18}$O structures in the entire G220 region have $\alpha \rm_{vir}<$2. 
{In addition, the median $\alpha \rm_{vir}$ values in the Perseus and Outer arms are much lower than those in the Local arm. This seems to be a consequence of the increasing detection limit. In the more distant arms, we miss \textbf{smaller} structures that tend to have relatively high $\alpha \rm_{vir}$ values (as shown in Section \ref{subsec:relations}).}


\subsubsection{Dense Gas Fraction}\label{subsubsec:ratios}
In the entire G220 region, there are 216 $^{12}$CO clouds ($ \sim $15\%) containing $^{13}$CO dense structures and 11 $^{12}$CO clouds ($ \sim $0.7\%) containing C$^{18}$O dense structures. The proportion of $ \sim $15\% is consistent with that measured by the MWISP survey in the SGQ \citep{Yuan_2022}. Besides, the proportions for $^{12}$CO clouds harboring $^{13}$CO dense structures are similar across the Galactic disk, with $ \sim $14\%, $ \sim $16\%, and $ \sim $14\% in the Local, Perseus, and Outer arms, respectively. One can see that typically only a small fraction of $^{12}$CO clouds harbor $^{13}$CO and/or C$^{18}$O dense structures. To understand the contributions of the dense structures to the MCs, we define the area ratios as follows: $\mathbf{A}^{13}_{12}\equiv A\left({\rm^{13}CO}\right)/A\left({\rm^{12}CO}\right)$ and $\mathbf{A}^{18}_{12}\equiv A\left({\rm C^{18}O}\right)/A\left({\rm^{12}CO}\right)$ where $A\left(\rm^{12}CO\right)$, $A\left(\rm^{13}CO\right)$, and $A\left(\rm C^{18}O\right)$ are the area of the $^{12}$CO cloud and the total areas of the $^{13}$CO and C$^{18}$O dense structures within each $^{12}$CO cloud, respectively. Similarly, we define the mass ratios as $\mathbf{M}^{13}_{12}\equiv M\left({\rm^{13}CO}\right)/M\left({\rm^{12}CO}\right)$ and $\mathbf{M}^{18}_{12}\equiv M\left({\rm C^{18}O}\right)/M\left({\rm^{12}CO}\right)$ where $M\left(\rm^{12}CO\right)$, $M\left(\rm^{13}CO\right)$, and $M\left(\rm C^{18}O\right)$ are the mass of the $^{12}$CO cloud and the total masses of the $^{13}$CO and C$^{18}$O dense structures within each $^{12}$CO cloud, respectively. 

The histograms of the ratios are plotted in Figure \ref{fig:ratio}. The area ratios $\mathbf{A}^{13}_{12}$ and $\mathbf{A}^{18}_{12}$ {span ranges from $\sim $1.2\% to 80\% and from $\sim $0.9\% to 5.9\%, respectively. The distributions of the mass ratios are similar to the area ratios, ranging from $ \sim $0.3\% to 76.6\% and from $ \sim$1.6\% to 12.3\% for $\mathbf{M}^{13}_{12}$ and $\mathbf{M}^{18}_{12}$, respectively.} \citet{Yuan_2022} reported a sharp upper limit of $\mathbf{A}^{13}_{12}=70\%$, and it is notable that, in general, our $\mathbf{A}^{13}_{12}$ values are indeed below this upper limit. {Moreover, we see that the median values (labeled in each panel) of $\mathbf{A}^{18}_{12}$ and $\mathbf{M}^{18}_{12}$ are quite small, suggesting that the C$^{18}$O dense structures typically occupy only a very small fraction of areas and masses in MCs.}
	
\subsection{Relations between Physical Properties}\label{subsec:relations}
	
Figure \ref{fig:relation} shows the relations between the physical properties of the $^{12}$CO (top panels), $^{13}$CO (center panels), and C$^{18}$O (bottom panels) structures. One can see that the $^{12}$CO clouds that contain $^{13}$CO and/or C$^{18}$O dense structures and the $^{13}$CO structures that contain C$^{18}$O dense structures generally have the largest sizes and masses and the most turbulent velocity dispersions. The log--log linear least-squares fitting results are {labeled} in the diagrams, with the Pearson correlation coefficients marked in parentheses. 
	
A power-law relation between mass and radius, $ M\propto R^2 $, is available from \citet{Larson_1981MNRAS}, indicating the clouds having a nearly constant surface density. \citet{Kauffmann_2010a,Kauffmann_2010b} {treated} the exponent of the $ M $--$ R $ relation as a proxy for how column density varies with radius. The left panels of Figure \ref{fig:relation} show the strong correlations (Pearson correlation coefficients $ r\gtrsim0.95 $) between {$ \log(M) $ and $ \log(R\rm_{eff}) $ for the} $^{12}$CO, $^{13}$CO, and C$^{18}$O structures, with the exponents $\beta=2.15$, 2.35, and 2.34, respectively. We find that the {$\beta$ values} fitted from the $^{13}$CO, and C$^{18}$O structures are very close to those measured by the Galactic Ring Survey (GRS) survey \citep{Roman-Duval_2010}, while those fitted from the $^{12}$CO structures are {closer} to Larson's results.
	
\begin{deluxetable*}{chcchhhhhhhhccchcccccchh}
	\tablecaption{{Basic Properties of Peculiar MCs}\label{tab:MCs}}
	\tablehead{
		\colhead{Name} & \nocolhead{Area} & \colhead{$ v\rm_{\scriptscriptstyle LSR} $} &
		\colhead{$ \sigma_v $}& \nocolhead{$ l $}& \nocolhead{$ l_{\rm rms} $}& \nocolhead{$ b $}& \nocolhead{$ b_{\rm rms} $}& \nocolhead{$ T_{\rm sum} $}& \nocolhead{$ N_{\rm pix} $}& \nocolhead{$ T_{\rm peak} $}& \nocolhead{$ N_{\rm channel} $}& \colhead{ Dist. }& \colhead{ $ R\rm_{eff} $}& \colhead{ Mass\tablenotemark{\footnotesize a} }& \nocolhead{$ M_{\rm vir} $}& \colhead{$ \alpha \rm_{vir} $}& \colhead{$\mathbf{A}^{13}_{12}$}& \colhead{$\mathbf{M}^{13}_{12}$}& \colhead{$\mathbf{A}^{18}_{12}$}& \colhead{$\mathbf{M}^{18}_{12}$}& \colhead{Arm}& \nocolhead{Flag}& \nocolhead{Matching}\\
		\colhead{} & \nocolhead{(arcmin$^2$)} & \colhead{($\rm km\,s^{-1}$)} &
		\colhead{($\rm km\,s^{-1}$)}& \nocolhead{($\degree$)}& \nocolhead{($\degree$)}& \nocolhead{($ \degree $)}& \nocolhead{($\degree $)}& \nocolhead{(K)}& \nocolhead{}& \nocolhead{(K)}& \nocolhead{}& \colhead{(kpc)}& \colhead{(pc)}& \colhead{($ M_\odot $)}& \nocolhead{($ M_\odot $)}& \colhead{}&\colhead{(\%)}&\colhead{(\%)}&\colhead{(\%)}&\colhead{(\%)}& \nocolhead{} & \nocolhead{$^{12}$CO cloud}
	}
	\decimalcolnumbers
	\startdata
	MWISP G220.634$-$1.916 &   3484.0 &   11.7 &   2.22 & 220.6 &   0.4 &  $-$1.9 &   0.3 &   566126.0 &  172982 &  27.5 &   82 &   1.0 &   9.7 &    $ 8.3\times10^3 $ &    55814.7 &     6.7 &52.8&42.7&1.8&4.0& Local & 12 & \nodata\\
	MWISP G221.775$-$3.007 &   4717.0 &   12.9 &   1.05 & 221.8 &   0.7 &  $-$3.0 &   0.4 &   681898.0 &  225857 &  22.2 &   57 &   1.0 &  11.3 &     $ 1.0\times10^4 $ &    14413.8 &     1.4 &44.5	& 37.7&0.9&1.6& Local & 12 & \nodata\\
	MWISP G224.440$-$1.069 &  21843.5 &   16.3 &   1.91 & 224.4 &   1.0 &  $-$1.1 &   0.7 &  5395960.9 & 1480675 &  26.4 &  107 &   1.1 &  27.9 &   $ 1.0\times10^5 $ &   118541.4 &     1.1 &50.5&52.7&3.9&9.1& Local & 12 & \nodata\\
	MWISP G221.921$-$2.121 &   1227.2 &   39.3 &   1.19 & 221.9 &   0.2 &  $-$2.1 &   0.1 &   318672.1 &   72359 &  25.4 &   66 &   3.7 &  21.4 &    $ 6.5\times10^4 $ &    35172.4 &     0.5 &58.6	&67.3&0.9&1.6& Perseus & 12 &\nodata\\
	MWISP G223.736$-$2.027 &    107.7 &   63.0 &   0.88 & 223.7 &   0.0 &  $-$2.0 &   0.1 &     7345.2 &    3712 &   6.0 &   36 &   6.6 &  11.2 &     $ 4.7\times10^3 $ &    10058.2 &     2.2 &22.3&18.8	&\nodata&\nodata& Outer & 12 & \nodata\\
	\enddata
	\tablenotetext{\footnotesize a}{Refers to the molecular mass traced by $^{12}$CO.}		
\end{deluxetable*}
	
\begin{figure*}[ht!]
	\gridline{\fig{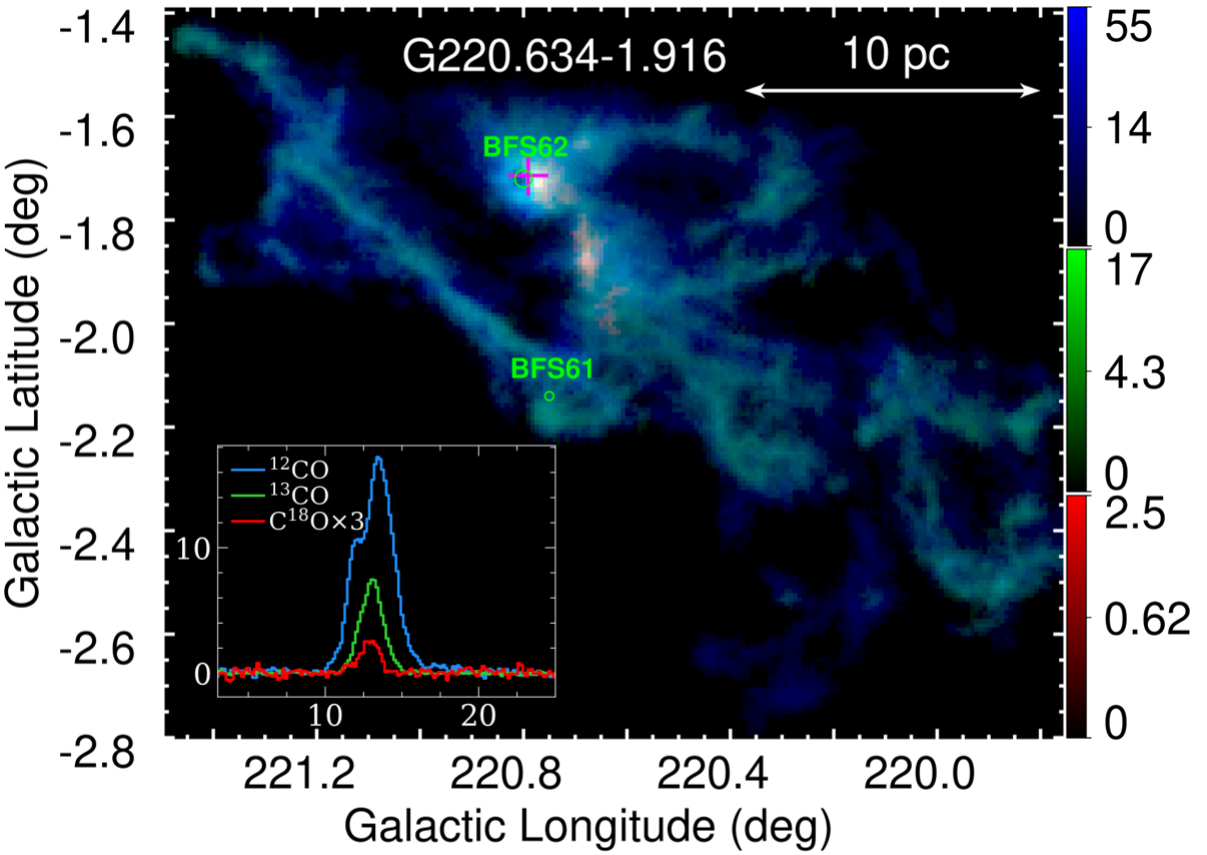}{0.48\textwidth}{}
		\fig{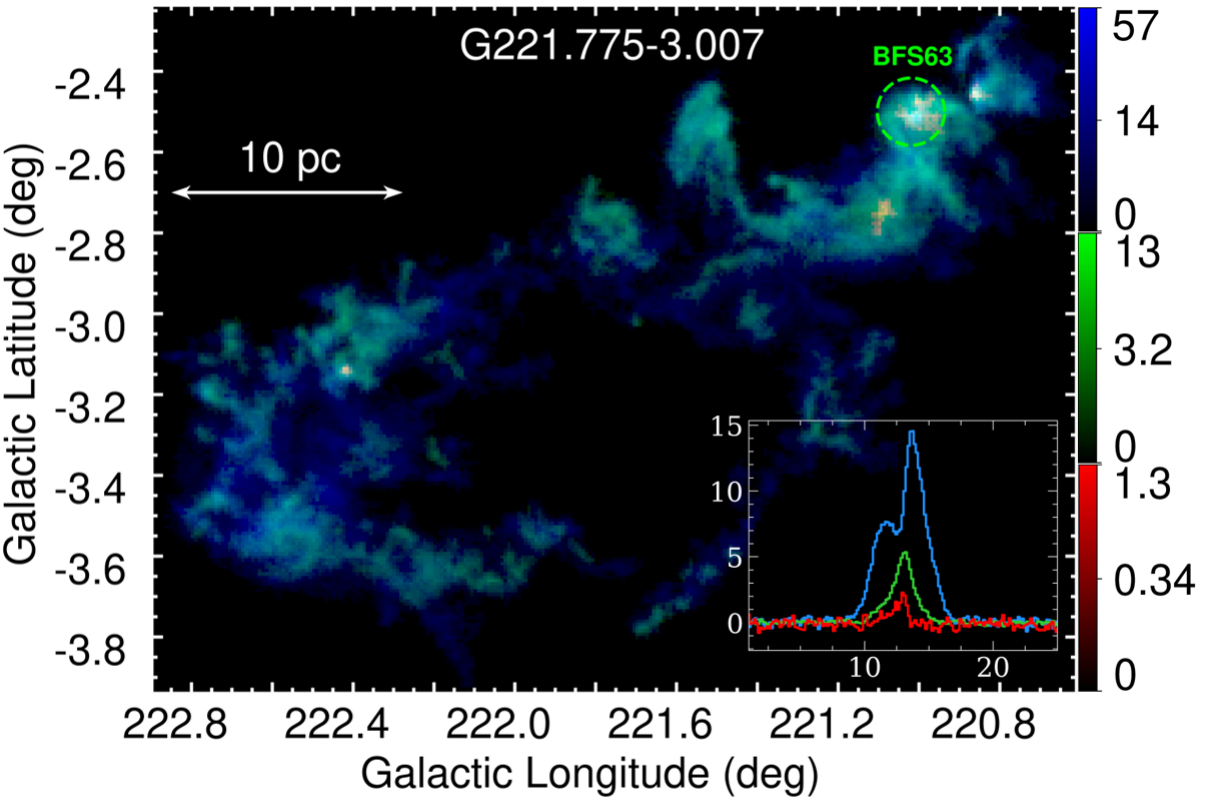}{0.52\textwidth}{}}
	\vspace{-0.3cm}
	\gridline{\fig{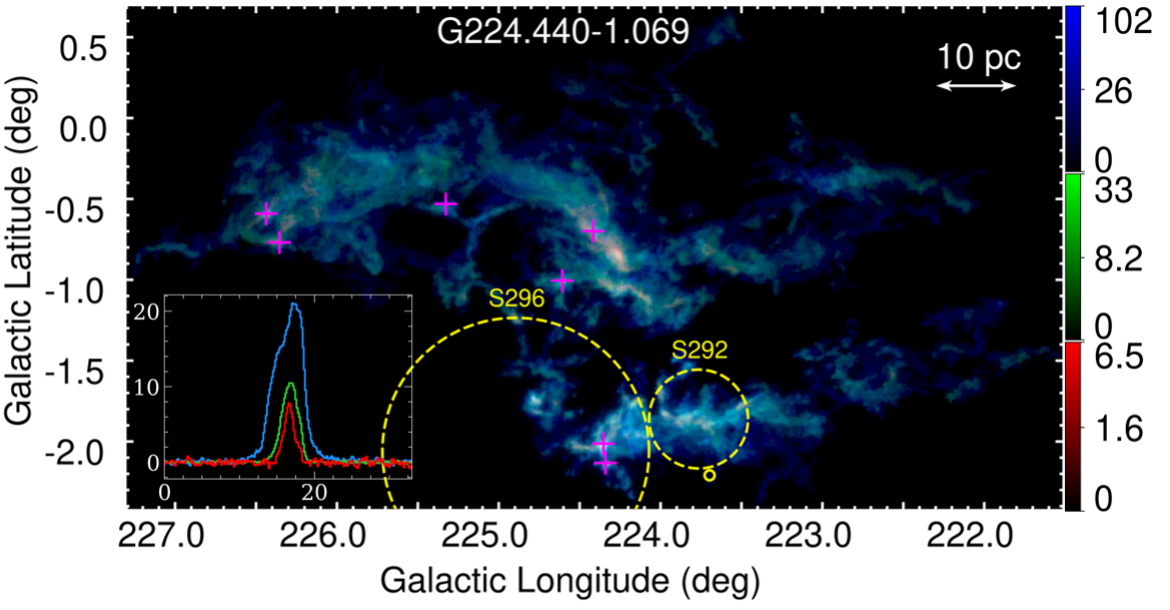}{0.67\textwidth}{}\fig{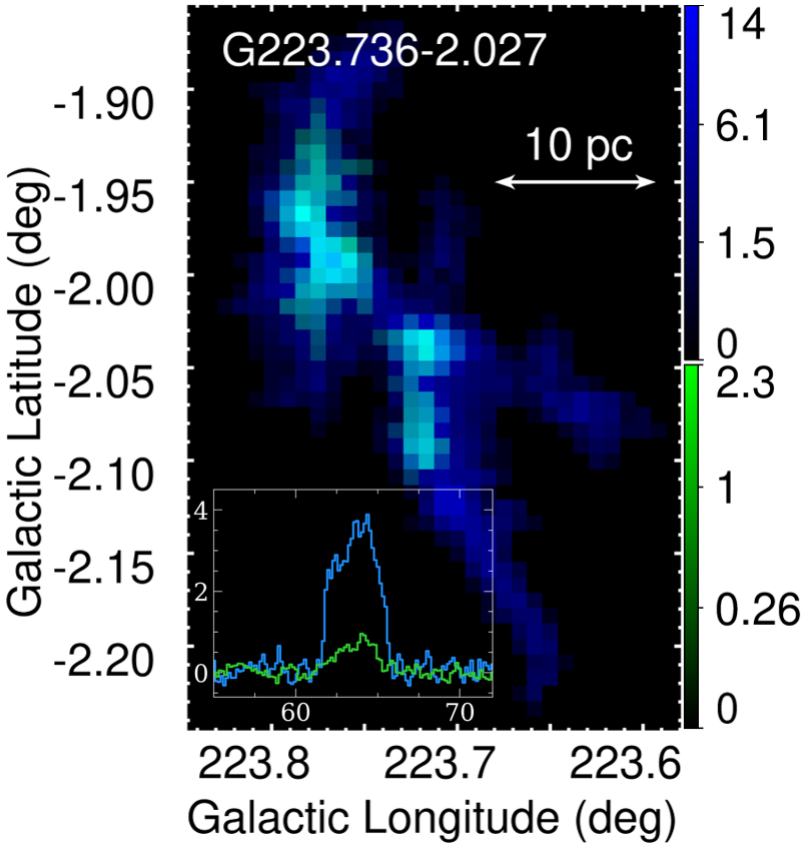}{0.33\textwidth}{}}
	\vspace{-0.3cm}
	\gridline{\fig{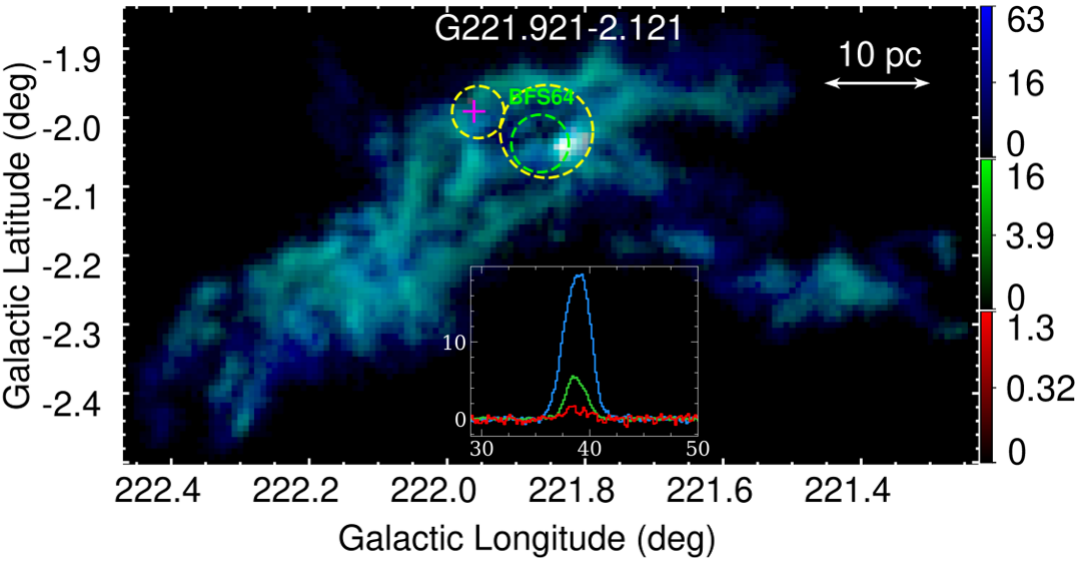}{0.65\textwidth}{}}
	\vspace{-0.4cm}
	\caption{Integrated intensity maps and {spectra of the }MCs MWISP G220.634$-$1.916, G221.775$-$3.007, G224.440$-$1.069, G223.736$-$2.027, and G221.921$-$2.121. The $^{12}$CO, $^{13}$CO, and C$^{18}$O lines are drawn in blue, green, and red, respectively. \textbf{The integrated intensity maps are shown in a square-root scale to clearly visualize the faint parts of the MCs,} and {the color bars represent integrated intensity in units of $\rm K~km\,s^{-1}$. The spectra shown in each plot\textbf{, with the y-axis on the $T_{\rm MB}$ scale,} are \textbf{averaged over} a $ 90\arcsec\times90\arcsec $ ($ 3\times3 $ pixels) region selected around the brightest pixel of each MC. The magenta crosses denote the 22 GHz H$_{2}$O masers detected toward the star formation regions \citep{Han_1998, Valdettaro_2001, Sunada_2007, Urquhart_2011}}. The \ion{H}{2} regions from \citet{Blitz_1982} and the WISE catalog {\citep{Anderson_2014}} are marked as green and yellow circles, respectively. \label{fig:MCs}}
\end{figure*}

The $ M $--$ R $ diagrams can also be used to investigate the possible high-mass star forming sites. The yellow shadings in Figure \ref{fig:relation} represent the region where clouds are not massive enough to form high-mass stars, i.e., $ M(r) \leqslant 870\,M_\odot\left(R/\rm pc\right)^{1.33} $ \citep{Kauffmann_2010a,Kauffmann_2010b, Kauffmann_2013,Kauffmann_2010c}. The solid red lines trace the empirical minimum threshold required for massive star formation---surface density equal to $ 0.05~\rm g\,cm^{-2} $ \citep{Urquhart_20131, Urquhart_20132}. We find that none of our structures reaches the threshold of \citet{Urquhart_20131, Urquhart_20132}, and only two $^{12}$CO clouds (MWISP G224.440$-$1.069 and MWISP G221.921$-$2.121; see Figure \ref{fig:MCs}) together with two $^{13}$CO structures (MWISP G223.939$-$1.895 and MWISP G221.954$-$2.090) lie above the border of \citet{Kauffmann_2010a,Kauffmann_2010b, Kauffmann_2013} and \citet{Kauffmann_2010c}. We also find that the two $^{13}$CO structures match the two $^{12}$CO clouds, respectively, which are the most massive MCs ($ \gtrsim$$6\times10^4~M_\odot $) in the G220 region. The scarcity of high-mass star formation is consistent with the literature, i.e., no 6.7 GHz methanol maser has been detected in the G220 region (according to the MaserDB database;\footnote{\url{https://maserdb.net}} \citealp{Ladeyschikov_2019}).
	
The velocity dispersion--size relation of the form $ \sigma_v\propto R^{0.5}$ is well known as the first Larson relation \citep[the cyan line in Figure \ref{fig:relation};][]{Larson_1981MNRAS, Solomon_1987,Heyer_2004}. The middle panels of Figure \ref{fig:relation} render the moderate correlations between $ \log(\sigma_v) $ and $ \log(R\rm_{eff}) $ with the Pearson correlation coefficients $ r=0.44$, 0.54, and 0.67 for the $^{12}$CO, $^{13}$CO, and C$^{18}$O structures, respectively. We find that {the exponents} traced by the $^{12}$CO and $^{13}$CO structures {in the TGQ} ($\beta=0.17 $ and 0.25, respectively) are {comparable with those observed by the MWISP survey in the SGQ \citep{Li_2020,Ma_2021}.} However, these {$\beta$ values} are much shallower than the Larson relation. {The shallow $ \sigma_v $--$ R\rm_{eff} $ relations may be due to the fact that the small clouds deviates from the Larson relation. Similar to \citet{Benedettini_2020}, we again perform least-squares fits to the $^{12}$CO clouds at two different scales. For those with radius $ R_{\rm eff}\geqslant2\rm~pc $, the best fit gives $\beta=0.55$ (and $ r=0.55 $), which is in good agreement with the Larson relation, whereas the best fit for those with $ R_{\rm eff}<2\rm~pc $ is flatter ($\beta=0.11 $ and $ r=0.25 $), which deviates significantly from the Larson relation. These are basically consistent with the results of \citet{Benedettini_2020}.} 
	
\begin{figure*}[ht!]
	\includegraphics[width=1\textwidth]{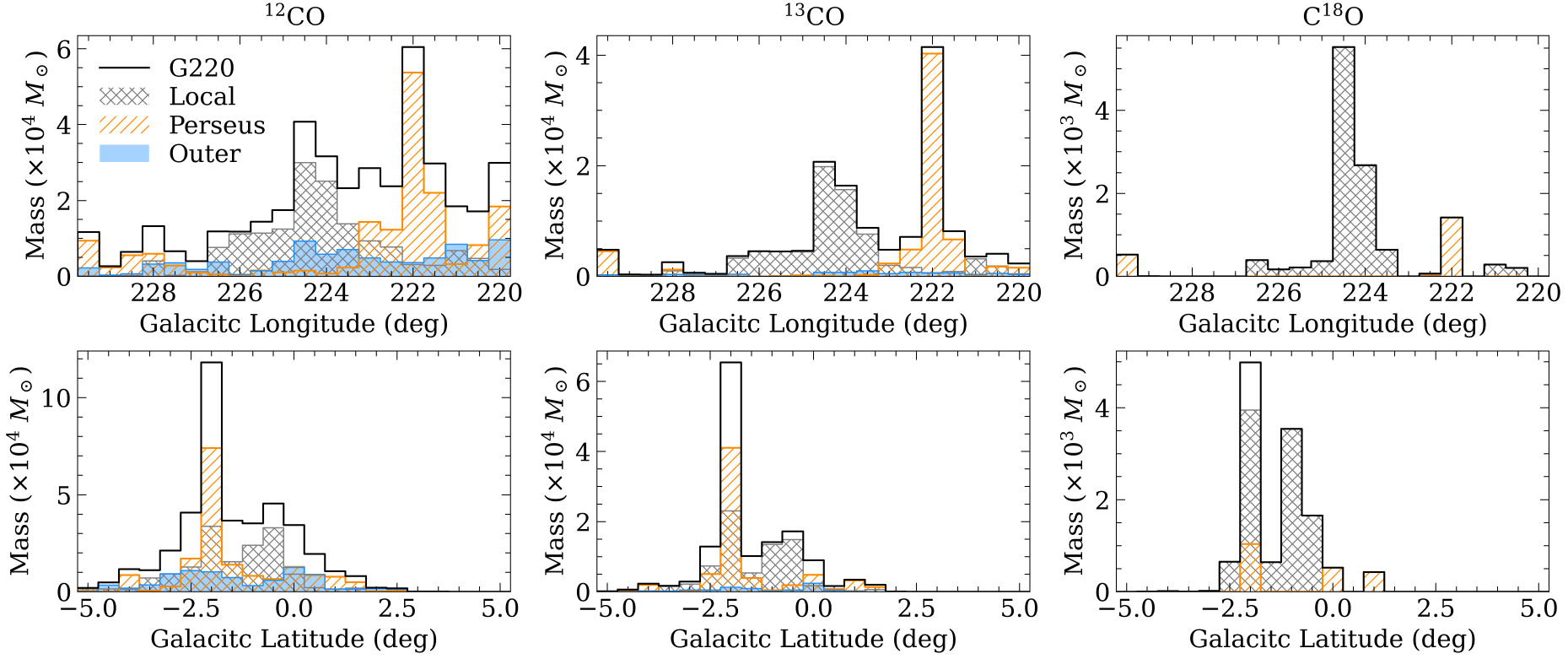}
	\caption{Molecular gas distributions along the Galactic longitude ({upper panels}) and the Galactic latitude ({lower panels}){, integrated over a Galactic latitude range from $-5\fdg25$ to $5\fdg25$ and a Galactic longitude range from $219\fdg75$ to $229\fdg75$, respectively. The mass distributions of the whole G220 region and the Local, Perseus, and Outer arms are shown in black, gray, orange, and blue histograms, respectively. And the} mass distributions traced by $^{12}$CO, $^{13}$CO, and C$^{18}$O are exhibited in the {left}, {middle}, and {right} panels, respectively. \label{fig:p-para}}
\end{figure*}
	
For pressure-confined objects (i.e., $ \alpha \rm_{vir}\gg1 $), \citet{Bertoldi_1992} {predicted} a theoretical power-law correlation between virial parameters and masses with an exponent equal to $ -\twothirds $. In Figure \ref{fig:relation} (right panels), anticorrelations over a broad mass range can be seen. We notice that the correlations for the $^{12}$CO and $^{13}$CO structures are well defined ($ r\sim-0.8$), while the correlation for C$^{18}$O structures is just modest ($ r=-0.39 $). The exponents for $^{12}$CO and $^{13}$CO structures ($=-0.36$ and $ -0.35 $, respectively) are similar to that reported by \citet{Kauffmann_2013} from their analysis of the GRS data \citep{Roman-Duval_2010}. These anticorrelation trends reflect a scenario in which more massive clouds are more gravitationally bound and therefore less stable---more likely to collapse without the additional support, such as from a strong magnetic field \citep{Kauffmann_2013}. Moreover, the masses in the bound $^{12}$CO, $^{13}$CO, and C$^{18}$O structures account for $ \sim $54.1\%, $ \sim $82.4\%, and $ \sim $96.8\% of the total $^{12}$CO, $^{13}$CO, and C$^{18}$O masses, respectively. These mass proportions for the bound structures are much higher than the number proportions (see Section \ref{subsubsec:clouds_para}). We can then conclude that, despite their small numbers, gravitationally bound molecular structures hold most of the molecular masses. 
	
\subsection{Massive MCs Presentation}\label{subsec:Example}
The high-sensitivity and high-resolution MWISP $^{12}$CO, $^{13}$CO, and C$^{18}$O data provide an excellent opportunity to observe the elaborate morphologies of the MCs. In this section, we present five {peculiar} MCs that {show filamentary structures and represent the most massive MCs in the three spiral arms.} Their basic properties are tabulated in Table \ref{tab:MCs}. {Interestingly, the dense gas fractions in these clouds are generally higher than the typical values marked in Figure \ref{fig:ratio} for each arm component.} Their integrated intensity maps and {spectra} are shown in Figure \ref{fig:MCs}, with $^{12}$CO, $^{13}$CO, and C$^{18}$O drawn in blue, green, and red, respectively. In each image, we also mark the associated 22 GHz H$_{2}$O maser sources, by using the MaserDB database \citep{Ladeyschikov_2019}, and the associated \ion{H}{2} regions, by referring to \citet{Blitz_1982} and the Wide-Field Infrared Survey Explorer (WISE) catalog of Galactic \ion{H}{2} regions v2.2\footnote{\url{http://astro.phys.wvu.edu/wise/}} \citep[hereafter, the WISE catalog]{Anderson_2014}. The following descriptions of the MCs are in the order of the maps in Figure \ref{fig:MCs}.
	
{MWISP G220.634$-$1.916}. This cloud has the largest velocity dispersion ($ \sim$$2.2\rm~km\,s^{-1}$) in the G220 region. {The optically thick $^{12}$CO spectral line around the emission peak exhibits an obvious red asymmetric structure.} 
{We find a that 22 GHz H$_{2}$O maser \citep{Han_1998} and two \ion{H}{2} regions \citep{Blitz_1982} are associated with this cloud}.
	
{MWISP G221.775$-$3.007}. The optically thick $^{12}$CO molecular line around the emission peak shows a significant red asymmetric structure. 

{MWISP G224.440$-$1.069}. This cloud is the most massive MC, accounting for $ \sim $26\%, $ \sim $38\%, and $ \sim $76\% of the total $^{12}$CO, $^{13}$CO, and C$^{18}$O traced masses in the entire G220 region, respectively. It traces the CMa OB1 complex, which has been studied in detail by the previous works, such as \citet{Kim_2004}, \citet{Elia_2013}, and \citet{Lin_2021}. {A total of seven 22 GHz H$_{2}$O masers (e.g., \citealp{Sunada_2007,Urquhart_2011}) and three \ion{H}{2} regions (according to the WISE catalog) are associated with this cloud.}

{MWISP G223.736$-$2.027}. It does not contain any C$^{18}$O dense structures. This source shows the typical filamentary structure even at a distance of $ \sim $6.6 kpc. {None of the known 22 GHz H$_{2}$O masers and \ion{H}{2} regions are associated with this cloud.} 

{MWISP G221.921$-$2.121}. This is the second most massive MC in the G220 region. A 22 GHz H$_{2}$O maser has been detected toward it (e.g., \citealp{Valdettaro_2001}). Two \ion{H}{2} regions from the WISE catalog and one \ion{H}{2} region from \citet{Blitz_1982} are associated with this cloud.

\begin{figure*}[ht!]
	\gridline{\fig{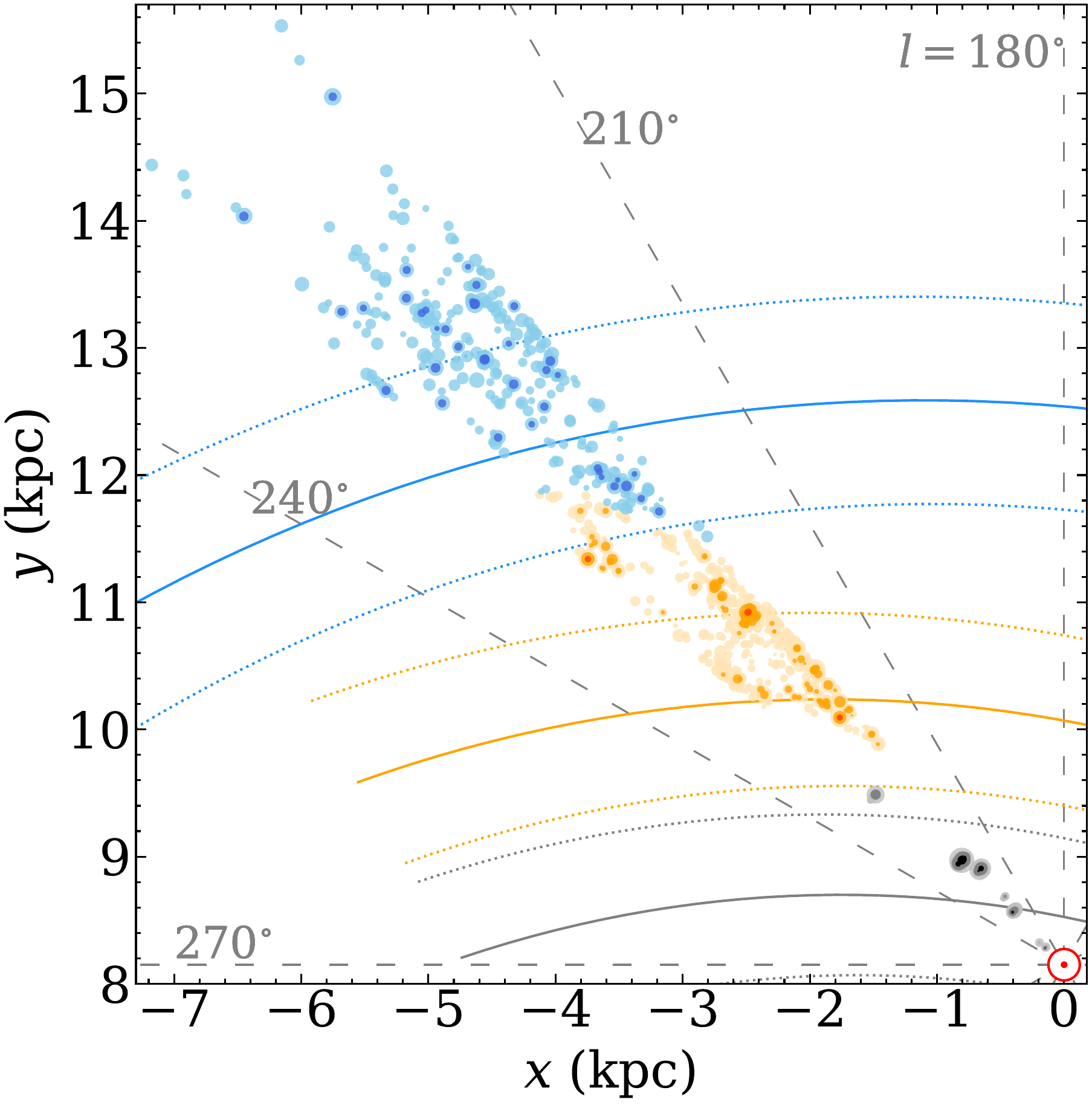}{0.412\textwidth}{}
		\fig{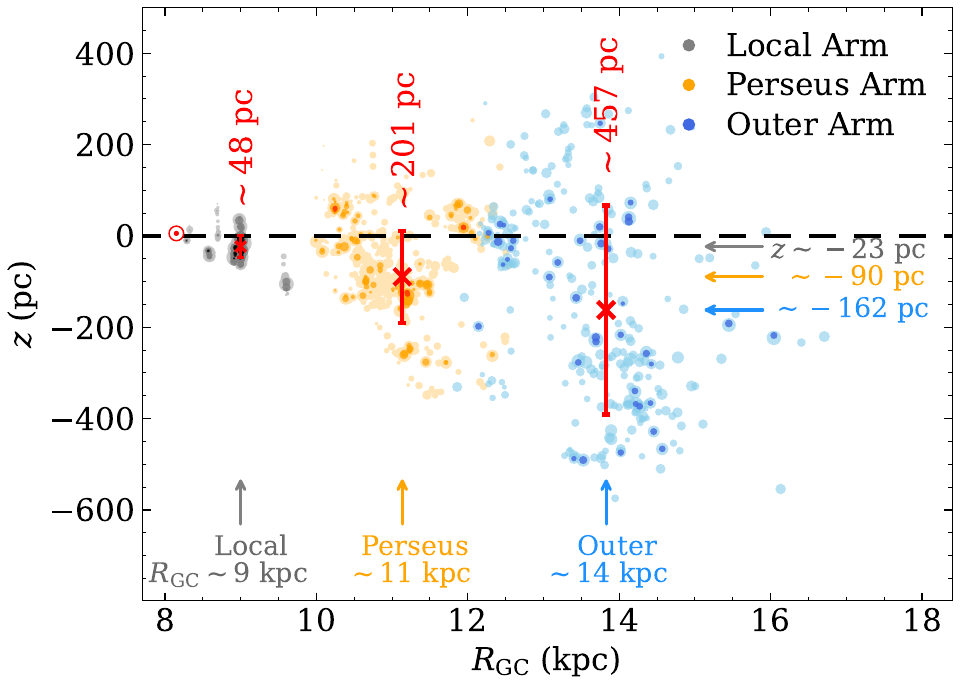}{0.588\textwidth}{}
	}
	\vspace{-0.8cm}
	\caption{{Left}: locations of the MCs projected on the Galactic plane. The plan view of the Milky Way \citep{Reid_2019} is overlaid. The straight dashed lines denote the directions of the Galactic longitude $l=180\degree$, $210\degree$, $240\degree$, and $270\degree$. The Sun (red Sun symbol) is at (0, 8.15) kpc. The solid gray, orange, and blue curved lines trace the centers (and the dotted curved lines indicate the widths enclosing $90\%$ of the high-mass star-forming regions of \citealt{Reid_2019}) of the Local, Perseus, and Outer arms, respectively. {Right}: distribution of the $\it z$-scale height vs. the Galactocentric radius $R\rm_{GC}$. The Galactic plane ($ z=0 $) is drawn with the dashed line. The Sun (red Sun symbol) is at ($8.15\rm~kpc$, $5.5\rm~pc$). The red crosses indicate the centers of $R\rm_{GC}$ and $\it z$ of the three arms, whose values are also marked. The red bars denote the thicknesses of the arms, with their values noted on the side. In both panels, the lightest, medium, and darkest filled circles represent the $^{12}$CO, $^{13}$CO, and C$^{18}$O structures, respectively. The symbol size is proportional to $\log\left(M\right)$. The proportions of the three CO isotopologs are different for the convenience of presentation. \label{fig:distribution}}
\end{figure*}

\section{3D Distributions in the Milky Way} \label{sec:gas_distribution}

In order to understand the mass distributions, we calculate the total molecular mass within every $0.5\degree$ of the Galactic longitude and latitude, respectively. The mass distributions along the $l$ (upper panels) and $b$ (lower panels) directions are depicted in Figure \ref{fig:p-para}, with those traced by $^{12}$CO, $^{13}$CO, and C$^{18}$O plotted in the left, center, and right panels, respectively. 

{We notice that in each individual spiral arm, the distribution peaks of the masses traced by $^{12}$CO, $^{13}$CO, and C$^{18}$O are approximately at the same $l$ and $b$. Along the $l$ direction, most of the molecular masses are distributed in the range of $l<225\degree$. The mass distributions in the Local arm are concentrated at $l\sim224\degree$, while those in the Perseus arm are concentrated at $l\sim222\degree$. And there is no significant mass peak along the $l$ direction in the Outer arm. This suggests that the molecular gas is unevenly distributed along the Local and Perseus arms, but appears to be more evenly distributed along the Outer arm. After close examination, we find that the distribution peaks of the Local and Perseus arms correspond to the aforementioned massive MCs MWISP G224.440$-$1.069 and MWISP G221.921$-$2.121~(Figure \ref{fig:MCs}), respectively. Along the $b$ direction, the majority of the molecular gases are distributed in the range of $b<0\degree$. According to the $^{13}$CO maps of \citet{Kim_2004}, the Local arm in the G220 region extends to at least $ b\sim-10\degree $, therefore it is important to note that our data do not actually cover the entire Local arm {(also mentioned in Section~\ref{subsec:large_scale})}. However, the Perseus and Outer arms seem to be almost completely mapped. The masses of the Perseus arm are concentrated at $b\sim-2\degree$. The Outer arm has no significant distribution peak along the $b$ direction either.}

{The face-on and vertical distributions of the MCs are displayed in Figure \ref{fig:distribution}, with the circle sizes proportional to $\log\left(M\right)$. In both maps, the Outer arm {traced by the molecular gas appears prominently} in this Galactic range for the first time, although it has been well described by other spiral arm tracers~\citep{Carraro_2005, Moitinho_2006, Vaquez_2008, Koo_2017}. 
Furthermore, we find that the spiral arms traced by MCs are different from the arm model of \citet{Reid_2019} in the face-on map. The discrepancy may be attributed to both the uncertainty of the kinematic distances and the bias of very few maser targets in the arm model. Along the $z$-direction, the MCs seem to be asymmetrically distributed. The minimum $z$ of the MCs reaches $\sim$$-600\rm~pc$, while the maximum $z$ only reaches $\sim$$400\rm~pc$. A substantial proportion ($ \sim$$75\% $) of the molecular structures fall below the Galactic plane. In addition, we see that the dynamic ranges of the MCs' $z$-scale heights are increasing with the Galactocentric radius $ R_{\rm GC}$. These reveal evident pictures of the Galactic disk warping and flaring {\citep{HI_Burke_1957,HI_Kerr_1957,Westerhout_1957, Burton1988, Wouterloot_1990}} in the G220 {region}.} 

{To better quantify the warp and flare of the molecular gas disk, we derive the arm ($ R_{\rm GC}$, $z$) centers by the mass-weighted average of the MCs' $ R_{\rm GC}$ and $z$-scale height, and we derive the arm thicknesses by the mass-weighted standard deviation of the MCs' $z$-scale height. The center positions are drawn as red crosses, and the thicknesses are marked with red bars in Figure~\ref{fig:distribution}. The derived ($ R_{\rm GC}$, $z$) centers of the Local, Perseus, and Outer arms are located at about (9, $-$23), (11, $-$90), and (14, $-$162) in units of (kpc, pc), respectively. The thicknesses of the Local, Perseus, and Outer arms are about 48, $201$, and $457$ pc, respectively. The Outer arm is more than twice as thick as the Perseus arm. Obviously, both the vertical displacements and thicknesses of the spiral arms are increasing with $ R_{\rm GC}$. These results are consistent with those traced by other spiral arm tracers in the same Galactic longitude range, such as \ion{H}{1} and stellar emission \citep[e.g.,][]{Gum_1960, Russeil_2003}.}

\section{Summary}\label{sec:sum}
Using the high-quality MWISP data, we conducted a large-scale $^{12}$CO, $^{13}$CO, and C$^{18}$O ($ J=1 $--0) survey in the G220 region, i.e., $l=[\,219\fdg75,~ 229\fdg75\,]$ and $b=[\,-5\fdg25,~5\fdg25\,]$ (105 deg$^2$ in total), and obtained 1,514,461 spectra for each CO isotopolog. According to the $ l $--$ v $ maps, the entire G220 region is divided into the Local, Perseus, and Outer arms that are discussed. Our main results can be summarized as follows.
\begin{enumerate}
\item We identified a total of 1,502 $^{12}$CO, 570 $^{13}$CO, and 53 C$^{18}$O molecular structures by using the DBSCAN algorithm. There are 216 $^{12}$CO structures containing $^{13}$CO dense structures and 11 $^{12}$CO structures containing C$^{18}$O dense structures.
	
\item The G220 region is generally in the low-excitation and low-density phase. The total masses of the $^{12}$CO, $^{13}$CO, and C$^{18}$O structures are estimated to be $4.0\times10^5 $, $ 1.4\times10^5 $, and $ 1.2\times10^4~M_\odot $, respectively. The Local, Perseus, and Outer arms account for $ \sim$$38\% $, $ \sim$$41\% $, and $ \sim$$21\% $ of the total $^{12}$CO-traced mass in the G220 region, respectively. Comparing to the SGQ \citep{Sun_2020}, we notice that in the TGQ, the Outer arm appears to be more conspicuous, while the Perseus arm is less prominent. 
	
\item Our survey resolves the molecular structures at different scales, i.e., with the effective radii of $ \sim $0.04--$ 27.9\rm~pc$ and masses of $ \sim$$10^{-2}$--$10^5~M_\odot $. The vast majority of the $^{12}$CO and $^{13}$CO structures appear to be gravitationally unbound ($\alpha \rm_{vir}\geqslant2$), while the C$^{18}$O structures are more likely to be bound ($\alpha \rm_{vir}<2$). 

\item Strong $ M $--$ R\rm_{eff} $ correlations and tight $ \alpha \rm_{vir} $--$ M $ anticorrelations are observable. However, the correlations between $ \sigma_v $ and $ R\rm_{eff} $ appear to be not obvious over the full dynamic range.
	
\item {For the Local, Perseus, and Outer arms, the derived arm $( R_{\rm GC},~z)$ centers are about (9, $-$23), (11, $-$90), and (14, $-$162) in units of (kpc, pc), and the derived arm thicknesses are about 48, $201$, and $457\rm~pc $, respectively. Both the vertical displacements and thicknesses of the spiral arms are increasing with the Galactocentric radius.}
	
\end{enumerate}

\begin{acknowledgments}
This research has made use of the data from the Milky Way Imaging Scroll Painting (MWISP) project, which is a multiline survey in $^{12}$CO/$^{13}$CO/C$^{18}$O along the northern Galactic plane with the PMO $13.7\rm~m$ telescope. We are very grateful to all members of the MWISP working group, particularly the staff members at the PMO $13.7\rm~m$ telescope, for their long-term support. MWISP was sponsored by the National Key R\&D Program of China with grant 2017YFA0402701 and the CAS Key Research Program of Frontier Sciences with grant QYZDJ-SSW-SLH047. Y.S. acknowledges support by the Youth Innovation Promotion Association, CAS (Y2022085), and the ``Light of West China'' Program (No. xbzg-zdsys-202212). J.Y. is supported by the National Natural Science Foundation of China through grant 12041305.
\end{acknowledgments}

\vspace{5mm}
\facility{PMO $13.7\rm~m$ telescope}

\software{GILDAS/CLASS \citep{GILDAS}, Astropy \citep{Astropy2013,Astropy2018}, matplotlib-v3.4.3\footnote{\url{https://doi.org/10.5281/zenodo.5194481}} \citep{Hunter_2007}, NumPy \citep{Numpy}}, SAOImage DS9 \citep{ds9}.

\appendix

\begin{splitdeluxetable*}{cccchhhhcrrBrrrrrrccc}
	\tabcolsep=2pt
	\tablenum{B1}
	\tabletypesize{\small}
	\tablecaption{Physical Properties of Molecular Structures\label{tab:catalog}}
	\tablewidth{700pt}
	\tablehead{
		\colhead{Index} & \colhead{Name} & \colhead{$ v\rm_{\scriptscriptstyle LSR} $} &
		\colhead{$ \sigma_v $}& \nocolhead{$ l $}& \nocolhead{$ l_{\rm rms} $}& \nocolhead{$ b $}& \nocolhead{$ b_{\rm rms} $}& \colhead{$ W\rm_{CO} $}&  \colhead{$ T_{\rm peak} $}&  \colhead{ Dist. }& \colhead{ $ R\rm_{eff} $ } & \colhead{Area}& \colhead{ Mass }& \colhead{$ \Sigma $}&\colhead{$ M_{\rm vir} $}& \colhead{$ \alpha \rm_{vir} $}& \colhead{Arm}& \colhead{Flag}& \colhead{Matching}\\
		\colhead{ } & \colhead{} & \colhead{($\rm km\,s^{-1}$)} &
		\colhead{($\rm km\,s^{-1}$)}& \nocolhead{($\degree$)}& \nocolhead{($\degree$)}& \nocolhead{($ \degree $)}& \nocolhead{($\degree $)}& \colhead{($\rm K~km\,s^{-1}$)}& \colhead{(K)}& \colhead{(kpc)}& \colhead{(pc)} & \colhead{(pc$^2$)}& \colhead{($ M_\odot $)}& \colhead{($ M_\odot\rm~pc^{-2}$)}& \colhead{($ M_\odot $)}& \colhead{}& \colhead{}& \colhead{} & \colhead{$^{12}$CO cloud}
		\decimalcolnumbers
	}
	\startdata
	4 & MWISP G220.634$-$1.916 &    11.66 &   2.22 & 220.634 & 0.380 & $-$1.916 & 0.258 & 6.4 &       27.5 &  1.00 &   9.7$ \quad $ & 294.8 $ \quad $&     $8.3\times10^3$ &       28.1  $ \quad $& $5.6\times10^4$ &   $\,$6.7$\,$ & Loc & 12 & \nodata    \\
	892 & MWISP G222.927$+$0.922 &    31.31 &   0.31 & 222.927 & 0.017 & 0.922 & 0.012 & 2.3 &        9.2 &  2.84 &   1.6$ \quad $ &   8.9 $ \quad $&       $8.7\times10^1$&        9.8  $ \quad $ &   $1.8\times10^2$ &   $\,$2.1$\,$ & Per & 12 & \nodata    \\
	1282 & MWISP G223.302$-$0.333 &    53.71 &   0.40 & 223.302 & 0.018 & $-$0.333 & 0.023 & 2.1 &        5.8 &  5.33 &   4.3$ \quad $ &  60.7 $ \quad $&      $5.6\times10^2$ &        9.2  $ \quad $&  $8.0\times10^2$ &   $\,$1.4$\,$ & Out & 12 & \nodata    \\
	1640 & MWISP G220.312$-$1.740 &    13.80 &   0.53 & 220.312 & 0.031 & $-$1.740 & 0.041 & 1.1 &        3.6 &  1.00 &   1.4$ \quad $ &   6.2 $ \quad $&       $8.8\times10^1$ &       14.2  $ \quad $&   $4.7\times10^2$&   $\,$5.3$\,$ & Loc & 13 & 4 \\
	1975 & MWISP G221.954$-$2.090 &    39.59 &   0.87 & 221.954 & 0.150 & $-$2.090 & 0.110 & 2.8 &        9.1 &  3.73 &  14.5$ \quad $ & 661.7 $ \quad $&    $4.1\times10^4$ &       61.2  $ \quad $ & $1.3\times10^4$ &   $\,$0.3$\,$ & Per & 13 & 953 \\
	2061 & MWISP G227.939$-$0.138 &    66.08 &   0.36 & 227.939 & 0.009 & $-$0.138 & 0.008 &  0.6 &        1.4 &  6.59 &   2.3$ \quad $ &  19.3 $ \quad $&      $1.6\times10^2$ &        8.4  $ \quad $&   $3.5\times10^2$ &   $\,$2.1$\,$ & Out & 13 & 1421 \\
	2068 & MWISP G229.744$+$0.117 &    70.29 &   0.33 & 229.744 & 0.008 & 0.117 & 0.020 &  0.6 &        1.8 &  6.99 &   3.0$ \quad $ &  30.0 $ \quad $&      $2.7\times10^2$ &        9.1  $ \quad $ &   $3.7\times10^2$ &   $\,$1.4$\,$ & Out & 13 & 1468 \\
	2077 & MWISP G220.677$-$1.866 &    12.28 &   0.40 & 220.677 & 0.012 & $-$1.866 & 0.028 &  0.5 &        2.2 &  1.00 &   0.8$ \quad $ &   1.9 $ \quad $&      $1.2\times10^2$ &       65.0  $ \quad $ &   $1.4\times10^2$ &   $\,$1.1$\,$ & Loc & 18 & 4 \\
	2125 & MWISP G229.578$+$0.153 &    52.90 &   0.60 & 229.578 & 0.011 & 0.153 & 0.005 &  0.5 &        1.1 &  4.92 &   1.5$ \quad $ &   8.2 $ \quad $&      $5.3\times10^2$ &       64.1  $ \quad $&  $6.3\times10^2$ &   $\,$1.2$\,$ & Per & 18 & 1204 \\	
	\enddata
	\tablecomments{Column (1): index of the source. Column (2): source name given by the MWISP project and the Galactic coordinates of the cloud centroid. Columns {(3)--(4)}: LSR velocity and velocity dispersion. {Column (5): average integrated intensity across the area of the structure.} Column (6): peak main-beam temperature. Columns {(7)--(13)}: heliocentric distance, effective radius, area, mass, {mass surface density}, virial mass, and virial parameter. Column ({14}): spiral arm layer. Column (15): flag---``12'', ``13'', and ``18'' denote samples traced {(and their physical properties estimated)} by the $^{12}$CO, $ ^{13} $CO, and C$^{18}$O lines, respectively. Column ({16}): index of the matching $ ^{12} $CO cloud for the $ ^{13} $CO and C$ ^{18} $O structures. Only a small portion of the catalog is shown here, and a machine-readable version of the full table {is published at Science Data Bank: \url{https://doi.org/10.57760/sciencedb.07765} \citep{Data}}.}
\end{splitdeluxetable*}

\section{Data with the bad channels}\label{sec:raw_data}
\begin{figure}[ht!]
	\centering
	\includegraphics[scale=0.6]{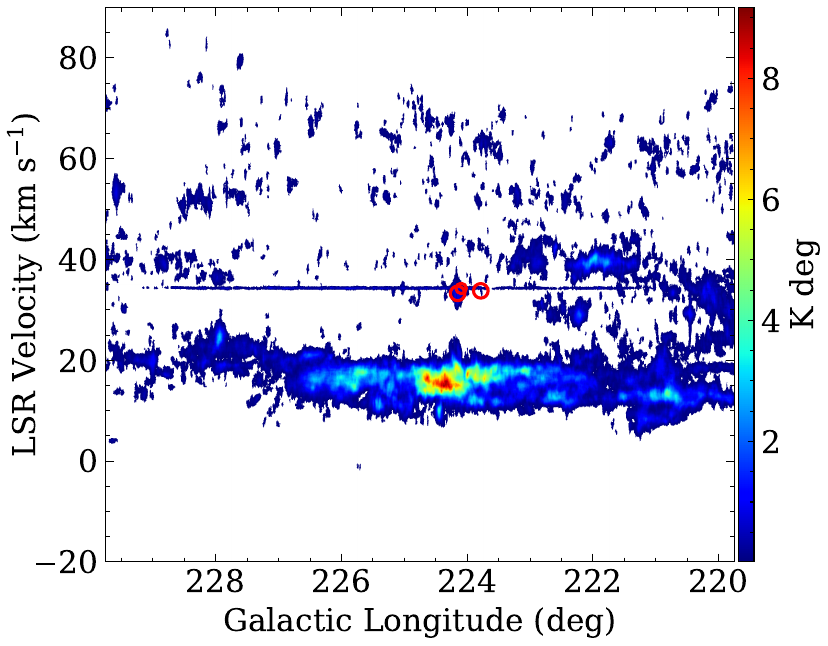}
	\figurenum{A1}
	\caption{Longitude--velocity map of the ``uncleaned'' $^{12}$CO emission{, integrated over a Galactic latitude range from $-5\fdg25$ to $5\fdg25$.} The three MCs in red circles are contaminated by the bad channels. \label{fig:raw_pvmap}}
\end{figure}

The $l$--$v$ map of the ``uncleaned'' $^{12}$CO emission is shown in Figure \ref{fig:raw_pvmap}. The integration ranges are $-5\fdg25\leqslant b\leqslant5\fdg25$. The map includes the spurious signals caused by $ \sim$$ 34~\rm km\,s^{-1} $ bad channels.

\section{Properties of Molecular Structures}\label{sec:catalog}
{The physical properties of the $^{12}$CO, $^{13}$CO, and C$^{18}$O molecular structures} are summarized in Table \ref{tab:catalog}.

\section{Derivation of Physical Properties}\label{sec:derivation}
{This appendix presents the equations used to estimate the physical properties. 
}
\subsection{Excitation Temperature}
Assuming that the $^{12}$CO line is optically thick and that the beam-filling factors are near unity, the excitation temperature in each pixel of the MC can be expressed as (e.g., \citealp{Bourke_1997})
\begin{equation}\label{equ:Tex}
	T_{\rm ex} = \frac{5.532}{\ln\left[\,1+5.532/\left(T\rm_{MB,pk}\!\left(^{12}CO\right)+0.819\right)\,\right]},
\end{equation}
where $T{\rm_{MB,pk}\!\left(^{12}CO\right)}$ is the peak main-beam brightness temperature of $^{12}$CO. 

\begin{figure*}[ht!]
	\centering
	\includegraphics[width=0.7\textwidth]{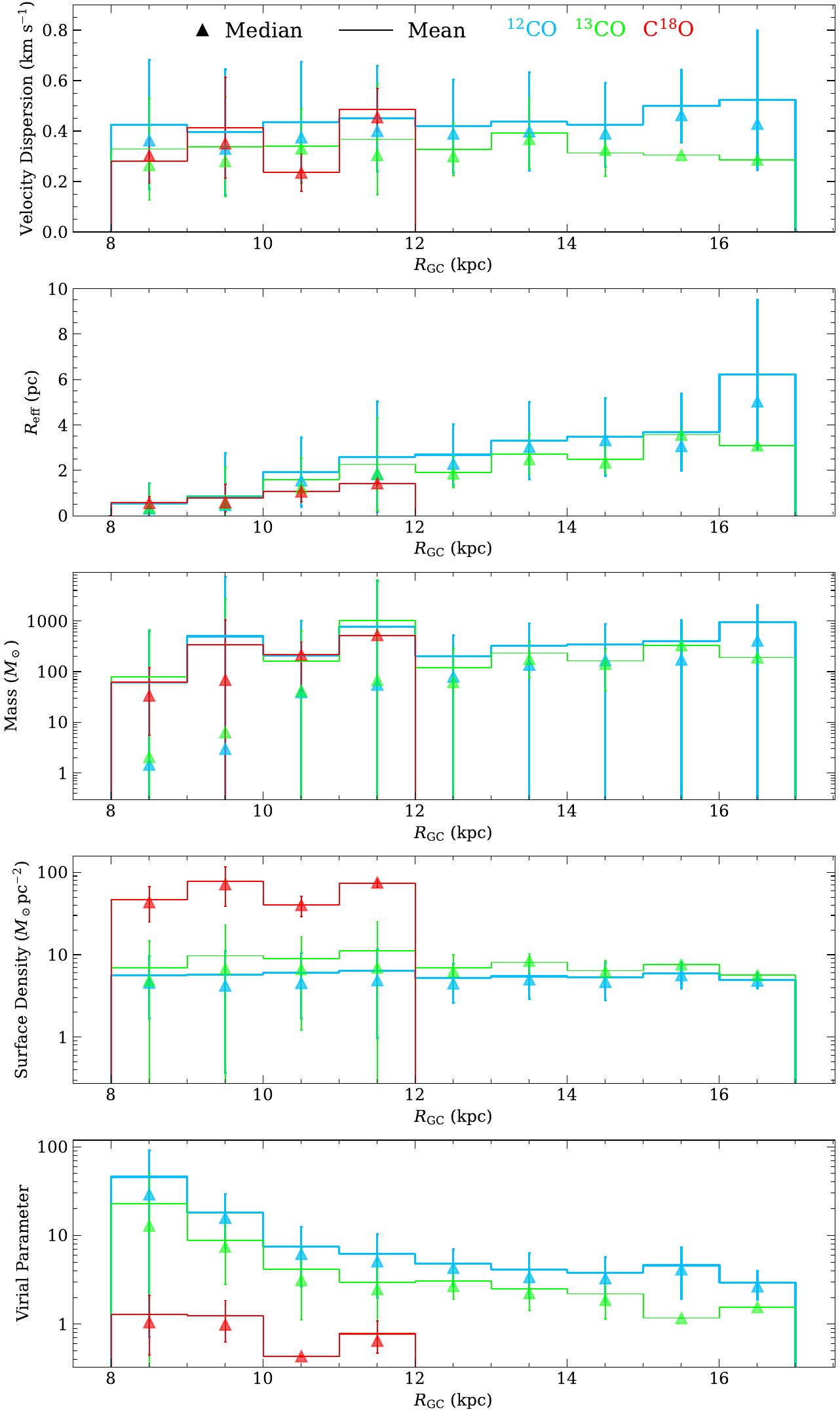}
	\figurenum{D1}
	\caption{{Physical properties of the $^{12}$CO (blue), $^{13}$CO (green), and C$^{18}$O (red) structures as a function of the Galactocentric radius. Histograms and triangles indicate the mean and median values of physical properties in the $ R\rm_{GC}$ bins of 1 kpc width, respectively, and vertical lines correspond to the standard deviations around the mean values. {From top to bottom}: velocity dispersion, effective radius, mass, mass surface density, and virial parameter. Note that the y-axis is logarithmically scaled for the physical properties whose mean or median values span more than $ \sim $2 orders of magnitude along $ R\rm_{GC}$.} \label{fig:vary}}
\end{figure*}

\subsection{$\rm H_2$ Column Density}
The H$_2$ column density {for the $^{12}$CO clouds} can be estimated by simply multiplying the integrated intensities of $^{12}$CO ($\int T{\rm_{MB}\!\left(^{12}CO\right)}\,dv$) by the commonly used CO-to-H$_2$ conversion factor $X_{\rm CO}=
2.0 \times 10^{20}~\rm cm^{-2}\left(K~km/s\right)^{-1}$ \citep{Bolatto_2013}. {For the $^{13}$CO and C$^{18}$O structures, the H$_2$ column density can be derived} using the LTE method, which is based on the assumption of the equal excitation temperatures for $^{12}$CO, $^{13}$CO, and C$^{18}$O in the LTE conditions. 

{Firstly, the optical depths of the $^{13}$CO and C$^{18}$O lines in each pixel of the molecular structure are given by}
\begin{align}\label{equ:tau13}
	\tau\!\left(\rm^{13}CO\right)=-\ln\!&\left[1-\frac{T{\rm_{MB,pk}\!\left(^{13}CO\right)}}{5.29}\right.\nonumber\\
	&\left.\times\left(\frac{1}{e^{5.29/T_{\rm ex}}-1}-0.164\right)^{-1}\right]
\end{align}
and
\begin{align}\label{equ:tau18}
	\tau\!\left(\rm C^{18}O\right)=-\ln\!&\left[1-\frac{T{\rm_{MB,pk}\!\left(C^{18}O\right)}}{5.27}\right.\nonumber\\
	&\left.\times\left(\frac{1}{e^{5.27/T_{\rm ex}}-1}-0.166\right)^{-1}\right],
\end{align}
where $T\rm_{MB,pk}\!\left(^{13}CO\right)$ and $T\rm_{MB,pk}\!\left(C^{18}O\right)$ are the peak main-beam brightness temperatures of $^{13}$CO and C$^{18}$O, respectively. 
{In general, the C$^{18}$O line is optically thin. For the $^{13}$CO line, we find that the median, 5th percentiles, and 95th percentile values of $\tau\!\left(\rm^{13}CO\right)$ are $ \sim $0.27, 0.12, and 0.67, respectively. This suggests that the $^{13}$CO line can be considered optically thin for almost all pixels of our $^{13}$CO structures. Then,} the $^{13}$CO and C$^{18}$O column densities in each pixel of {the molecular structure} can thus be expressed as \citep{Bourke_1997, Pineda_2010}
\begin{align}\label{equ:I_13}
	N_{\rm^{13}CO}=2.&42\times{10}^{14}\times\frac{\tau\!\left(\rm^{13}CO\right)}{1-e^{-\tau\!\left(\rm^{13}CO\right)}}\nonumber\\
	&\times\frac{1+0.88/T_{\rm ex}}{1-e^{-5.29/T_{\rm ex}}}	\int\! T_{\rm MB}\!\left(\rm^{13}CO\right)\,dv
\end{align}
and
\begin{align}\label{equ:I_18}
	N_{\rm C^{18}O}=2.&54\times{10}^{14}\times\frac{\tau\!\left(\rm C^{18}O\right)}{1-e^{-\tau\!\left(\rm C^{18}O\right)}}\nonumber\\
	&\times\frac{1+0.88/T_{\rm ex}}{1-e^{-5.27/T_{\rm ex}}}\int\! T_{\rm MB}\!\left(\rm C^{18}O\right)\,dv.
\end{align}

Finally, the isotope ratios of $\left[\rm^{12}C/^{13}C\right]=6.21\it R\rm_{GC}+18.71$ \citep[$R_{\rm GC}$ is the Galactocentric radius in units of kpc;][]{Milam_2005} and $\left[\rm^{16}O/^{18}O\right]=560$ \citep{Wilson_1994} and an abundance ratio of $\left[\rm H_2/^{12}C\right]=1.1\times10^4$ \citep{Frerking_1982} are adopted to convert the $^{13}$CO and C$^{18}$O column densities to H$_2$ column densities. 

\subsection{Mass, Mass Surface Density, Effective Radius, Velocity Dispersion, and Virial Parameter}
The mass of the molecular structure is estimated as $ M=2\mu m_{\scriptscriptstyle\rm H} a^2d^2\sum\nolimits_{i} N_{\rm H_2} $, where $\mu=1.36$ is the mean atomic weight, taking the contribution of helium and metals into account, $m\scriptscriptstyle\rm_H$ is the mass of a hydrogen {atom}, $a=30\arcsec$ is the angular size of a pixel, and $d$ is the heliocentric distance. 

The mass surface density, in units of $M_\odot~\rm pc^{-2}$, is simply calculated by $\Sigma=M/\left(A\times d^2\right)$, where $ A = a^{\rm2}\times N_{\rm pixel} $ is the angular area (projected area) of the molecular structure.

Assuming that the cloud is spherical, the effective radius can thus be expressed as $R_{\rm eff} = d\,\sqrt{\frac{A}{\pi}-\frac{\theta^2_{\rm MB}}{4}}$ \citep{Ladd_1994}, where $ \theta_{\rm MB}\sim 50\arcsec $ is the beam size.

The velocity dispersion ($\sigma_v$) of the molecular structure is defined as the $ T\rm_{MB} $-weighted second moment within the PPV voxels of the structure. Then, the line width of the molecular structure is estimated as $\Delta v=\sqrt{8\ln 2}\times \sigma_v$. 

The virial mass can be calculated by $ M_{\rm vir} = 210\times R_{\rm eff}\times\left(\Delta v\right)^2 $ \citep{MacLaren_1988}, where $R_{\rm eff}$ is in units of pc, $\Delta v$ is
in units of $\rm km\,s^{-1}$, and the derived $M_{\rm vir}$ is in units of $M_\odot$. Then, the virial parameter is defined as $\alpha_{\rm vir}=M_{\rm vir}/ M$.

\section{Physical Properties at different Galactocentric radii}\label{sec:vary}
{This appendix presents the mean and median values of physical properties as a function of the Galactocentric radius.
}

\bibliography{main_V2}{}
\bibliographystyle{aasjournal}

\end{document}